\documentclass[12pt]{article}
\usepackage{natbib,graphicx,amsfonts,amsmath,amssymb,mathrsfs, comment}

\title{On the Common Structure of Bohmian Mechanics and the
       Ghirardi--Rimini--Weber Theory}
   
\author{
Valia Allori\footnote{Department of Philosophy, Davison Hall,
     Rutgers, The State University of New Jersey, 26 Nichol Avenue,
     New Brunswick, NJ 08901-1411, USA.
     E-mail: vallori@eden.rutgers.edu},
Sheldon Goldstein\footnote{Departments of Mathematics, Physics and
     Philosophy, Hill Center, Rutgers, The State University of New  
     Jersey, 110 Frelinghuysen Road, Piscataway, NJ 08854-8019, USA.
     E-mail: oldstein@math.rutgers.edu},\\
Roderich Tumulka\footnote{Mathematisches Institut,
     Eberhard-Karls-Universit\"at, Auf der Morgenstelle 10, 72076
     T\"ubingen, Germany. E-mail:
     tumulka@everest.mathematik.uni-tuebingen.de},
 and Nino Zangh\`\i\footnote{Dipartimento di Fisica dell'Universit\`a
     di Genova and INFN sezione di Genova, Via Dodecaneso 33, 16146
     Genova, Italy. E-mail: zanghi@ge.infn.it}
}
\date{June 2, 2007}

\newcommand{\conf}{\mathcal{Q}}
\newcommand{\Q}{\conf}

\renewcommand{\Im}{\mathrm{Im}}
\newcommand{\EEE}{\mathbb{E}}

\newcommand{\PPP}{\mathbb{P}}
\newcommand{\RRR}{\mathbb{R}}
\newcommand{\CCC}{\mathbb{C}}

\newcommand{\NNN}{\mathbb{N}}

\renewcommand{\sp}[2]{\langle #1|#2 \rangle}

\newcommand{\macroX}{\mathscr{X}}

\newcommand{\wf}{wave function }
\newcommand{\po}{primitive ontology }

\begin{document}
\maketitle
\begin{abstract}
Bohmian mechanics and the Ghirardi--Rimini--Weber theory provide opposite resolutions of the quantum measurement problem:  the former postulates additional variables (the particle positions) besides the wave function, whereas the latter implements spontaneous collapses of the wave function by a nonlinear and stochastic modification of Schr\"odinger's equation. Still, both theories, when understood appropriately, share the following structure: They are ultimately not about wave functions but about `matter' moving in space, represented by either particle trajectories, fields on space-time, or a discrete set of space-time points. The role of the wave function then is to govern the motion of the matter.
\medskip

\end{abstract}

\begin{center}
\textit{Dedicated to GianCarlo Ghirardi on the occasion of his 70th birthday}  
\end{center}
\tableofcontents

\section{Introduction}

Bohmian mechanics ({\sf BM}) and the Ghirardi--Rimini--Weber ({\sf GRW})
theory are two quantum theories without observers, and thus provide two
possible solutions of the measurement problem of quantum
mechanics. However, they would seem to have little in common beyond
achieving the goal of describing a possible reality in which observers
would find, for the outcomes of their experiments, the probabilities
prescribed by the quantum formalism. They are two precise, unambiguous
fundamental physical theories that describe and explain the world around
us, but they appear to do this by employing opposite strategies. In Bohmian
mechanics \citep{Bohm52, Bell66, DGZ92, survey} the wave function evolves
according to the Schr\"odinger equation but is not the complete description
of the state at a given time; this description involves further variables,
traditionally called `hidden variables,' namely the particle
positions. In the {\sf GRW} theory \citep{Pe76, GRW86, Bell87, BG03}, in
contrast, the wave function $\psi$ describes the state of any physical
system completely, but $\psi$ collapses spontaneously, thus departing from
the Schr\"odinger evolution.  That is, the two theories choose different
horns of the alternative that Bell formulated as his conclusion from the
measurement problem \citep{Bell87}: `Either the wave function, as given by
the Schr\"odinger equation, is not everything, or it is not right.'

The two theories are always presented almost as dichotomical, as in the recent paper by \citet{putnam}.
Our suggestion here is instead that {\sf BM} and {\sf GRW} theory have much more in common than one would expect at first sight. 
So much, indeed, that they should be regarded as being close to each other, rather than opposite. 
The differences are less profound than the similarities, provided that the {\sf GRW} theory is understood appropriately, as involving variables describing matter in space-time.
These variables we call the \textit{primitive ontology} (PO) of the theory, and they form the common structure of {\sf BM} and {\sf GRW}. The gain from the comparison with {\sf BM} is the insight that the {\sf GRW} theory can, and should, be understood in terms of the PO. We think this view in terms of the PO provides a deeper understanding of the {\sf GRW} theory in particular, and of quantum theories without observer in general. To formulate more clearly and advertise this view is our goal.

After recalling what Bohmian mechanics is in Section \ref{sec:bm}, we introduce
two concrete examples of {\sf GRW} theories in Section \ref{sec:GRW}. These examples involve rather different choices of crucial variables, describing matter in space-time, and give us a sense of the range of possibilities for such variables. 
We discuss in Section \ref{sec:PO}   
the notion of the {\it{primitive ontology} }(PO) of a theory 
(a notion introduced in \citep{DGZ92})
and connect it to  
Bell's notion  of `local beables'   \citep{Bell76}.
In Section  \ref{sec:morePO} we relate the primitive ontology of a theory to the notion of physical equivalence between theories.   We stress in Section \ref{sec:symmetry} the connection, first discussed in \citep{Gol98}, between the \po and symmetry properties, with particular concern for the generalization of such theories to a relativistically invariant quantum theory without observers. In  Section \ref{sec:grw0} 
we argue that a theory without a primitive ontology is at best profoundly problematical.
We proceed in Sections \ref{sec:diff} to an analysis of the differences between {\sf GRW} (with primitive ontology) and {\sf BM}, and in Section \ref{sec:differences} we discuss a variety of possible theories. We consider in Section \ref{sec:lgrwf} a `no-collapse' reformulation of one of the {\sf GRW} theories and in Section \ref{sec:BMC} a `collapse' interpretation of {\sf BM}. These formulations  enable us to better appreciate the common structure of {\sf BM} and the {\sf GRW} theories, as well as the differences, as we discuss in Section \ref{sec:equiv}. We conclude in Section \ref{sec:essential} with a summary of this common structure.

\section{Bohmian Mechanics}
\label{sec:bm}

Bohmian mechanics is a theory of (nonrelativistic) particles in motion. The motion of a system of $N$ particles is provided by their world lines $t \mapsto Q_i(t)$, $i=1, \ldots, N$, where $Q_i(t)$ denotes the position in $\RRR^3$ of the $i$-th particle at time $t$. These world lines are determined by Bohm's law of motion \citep{Bohm52, Bell66, DGZ92, survey}, 
\begin{equation}\label{Bohm}
  \frac{dQ_i}{dt}=v_i^{\psi}(Q_1, \ldots, Q_N)=\frac{\hbar}{m_i}
  \Im \frac{\psi^{*}\nabla_i \psi}{\psi^{*}\psi}(Q_1\ldots,Q_N), 
\end{equation}
where $m_i$, $i=1, \ldots, N$,  are the masses of the particles;
the wave function $\psi$ evolves according to Schr\"odinger's equation
\begin{equation}\label{Schr}
 i\hbar\frac{\partial \psi}{\partial t} = H\psi \,,
\end{equation} 
where
 $H$ is the usual nonrelativistic Schr\"odinger Hamiltonian; for
spinless particles it is of the form
\begin{equation}
\label{eq:H}
H=-\sum_{k=1}^N\frac{\hbar^2}{2m_k}\nabla^2_k+V,
\end{equation}
containing as parameters the masses  of the particles as
well as the potential energy function $V$ of the system. 

In the usual yet unfortunate terminology, the actual positions $Q_1, ..., Q_N $ of the particles are the \emph{hidden variables} of the theory: the variables which, together with the wave function, provide a complete description of the system, the wave function alone providing only a partial, incomplete, description.
From the point of view of {\sf BM}, however, this is a strange terminology since it suggests that the main object of the theory is the wave function, with the additional information provided by the particles' positions playing a secondary role. The situation is rather much the opposite: {\sf BM} is a theory of particles; their positions are the primary variables, and the description in terms of them must be completed by specifying the wave function to define the dynamics \eqref{Bohm}. 

As a consequence of Schr\"odinger's equation and of Bohm's law of motion, 
the quantum equilibrium distribution $|\psi(q)|^2$ is equivariant.  This means that if the configuration $Q(t) = (Q_1(t), \ldots, Q_N(t))$ of a system is random with distribution $|\psi_t|^2$ at some time $t$, then this will be true also for any other time $t$. Thus, the \emph{quantum equilibrium hypothesis}, which asserts that whenever a system has wave function $\psi_t$, its configuration $Q(t)$ is random with distribution $|\psi_t|^2$, can consistently be assumed. This hypothesis is not as hypothetical as its name may suggest: the quantum equilibrium hypothesis follows, in fact, by the law of large numbers from the assumption that the (initial) configuration of the universe is typical (i.e., not-too-special) for the $|\Psi|^2$ distribution, with $\Psi$ the (initial) wave function of the universe \citep{DGZ92}. The situation resembles the way Maxwell's distribution for velocities in a classical  
gas follows from the assumption that the phase point of the gas is typical for the uniform distribution on the energy surface.

As a consequence of the quantum equilibrium hypothesis, a Bohmian universe, even if deterministic, appears random to its inhabitants. In fact, the probability distributions observed by the inhabitants agree exactly with those of the quantum formalism. To begin to understand why, note that any measurement apparatus must also consist of Bohmian particles. Calling $Q_{S}$ the configuration of the particles of the system to be measured and $Q_{A}$ the configuration of 
the particles of the apparatus,
we can write for the configuration of the big Bohmian system relevant to the analysis of the measurement $Q=(Q_{S},Q_{A})$.  Let us suppose that the initial wave function $\psi$ of the big system is a product state  $\Psi(q)=\Psi(q_{S},q_{A})= \psi(q_{S}) \, \phi(q_{A})$. 

During the measurement, this $\Psi$ evolves according to the Schr\"odinger equation, and in the case of an ideal measurement it evolves to $\Psi_t = \sum_\alpha \psi_\alpha \, \phi_\alpha$, where $\alpha$ runs through the eigenvalues of the observable measured, $\phi_\alpha$ is a state of the apparatus in which the pointer points to the value $\alpha$, and $\psi_\alpha$ is the projection of $\psi$ to the appropriate eigenspace of the observable. By the quantum equilibrium hypothesis, the probability for the random apparatus configuration $Q_A(t)$ to be such as to correspond to the pointer pointing to the value $\alpha$ is $\|\psi_\alpha\|^2$. For a more detailed discussion see \citep{DGZ92,DGZ04}. 

\section{Ghirardi, Rimini, and Weber}\label{sec:GRW}

The  theory  proposed by \citet*{GRW86} is in agreement with the predictions of nonrelativistic quantum mechanics as far as 
all present 
experiments are concerned \citep{BG03}; for a discussion of future experiments that may
distinguish this theory from quantum mechanics, see
Section~V of \citep{BG03}. According to the way in which this  theory is usually presented, the evolution of the wave function follows, instead of Schr\"odinger's equation, a stochastic jump process in Hilbert space. 
We shall succinctly summarize this process as follows.

Consider a quantum system described (in the standard language) by 
an $N$-`particle'\footnote{We wish to emphasize here that there are no  particles in this theory: the word `particle' is used only for convenience in order to be able to use the standard notation and terminology.} wave function $\psi = \psi(q_1,...,q_N)$, ${q}_i\in \RRR^3$, $i=1,\dots, N$;  for any point $x$ in $\mathbb{R}^3$ (the `center' of the collapse that will be defined next), define on the Hilbert space of the system  the \emph{collapse operator}
\begin{equation}
\Lambda_i (x) =\frac{1}{(2\pi \sigma^2)^{3/2}}\, e^{-\frac{( \widehat{Q}_i-x)^2}{2\sigma^2}}\,,
\label{eq:collapseoperator}
\end{equation}
where  $\widehat{Q}_i$ is the position operator of  `particle' $i$. Here $\sigma$ is a new constant of nature of order of $10^{-7}$m.

Let $\psi_{t_0}$ be the initial wave function, i.e., the normalized wave function at some time $t_0$ arbitrarily chosen as initial time. Then $\psi$ evolves in the following way: 
\begin{enumerate}
\item It evolves unitarily, according to Schr\"odinger's equation, until a random time $T_1= t_0 + \Delta T_1$, so that
\begin{equation}
\psi_{T_1}= U_{\Delta T_1} \psi_{t_0},
\end{equation}
where $U_t$ is the unitary operator $U_t=e^{-\frac{i}{\hbar}Ht}$ corresponding to the standard Hamiltonian $H$ governing the system, e.g., given by (\ref{eq:H}) for $N$ spinless particles, 
and  $\Delta T_1$ is a random time distributed according to the exponential distribution with rate $N\lambda$ (where the quantity $\lambda$ is another constant of nature of the theory,\footnote{\citet{PS94} have argued that $\lambda$ should be chosen differently for every `particle,' with $\lambda_i$ proportional to the mass $m_i$.} of order of $10^{-15}$ s$^{-1}$).
\item At time $T_1$ it undergoes an instantaneous collapse with random center 
 $X_1$ and random label $I_1$ according to
\begin{equation}
\psi_{T_1} \mapsto\psi_{T_1+}= \frac{\Lambda_{I_1} (X_{1})^{1/2}\psi_{T_1}}{\| \Lambda_{I_1} (X_{1})^{1/2} \psi_{T_1} \|}.
\end{equation}
$I_1$ is chosen at random in the set $\{1, \ldots, N\}$ with uniform distribution. The center  of the collapse 
$X_1$ is chosen randomly with probability distribution\footnote{Hereafter, when no ambiguity could arise, we use the standard notations of probability theory, according to which a capital letter, such as $X$, is used to denote a random variable, while the the values taken by it are denoted by small letters; $ X\in dx$ is a shorthand for $X\in [x, x+dx]$, etc. }
\begin{equation}\label{p}
\mathbb{P}(X_1\in dx_{1}|  \psi_{T_1}, I_1=i_1) = \left\langle \psi_{T_1}|\Lambda_{i_1}(x_1)\psi_{T_1}\right\rangle dx_{1}=\|\Lambda_{i_1} (x_1)^{1/2} \psi_{T_1}\|^2 dx_{1}.
\end{equation}
\item Then the algorithm is iterated: $\psi_{T_1+}$ evolves unitarily until a random time $T_2 =  T_1 + \Delta T_2$, where  $\Delta T_2$ is a random time (independent of $\Delta T_1$) distributed according to the exponential distribution with rate $N\lambda$, and so on.
\end{enumerate}

In other words, the evolution of the wave function is the Schr\"odinger evolution interrupted by collapses. When the wave function is 
$\psi$ a collapse with center $x$ and label $i$ occurs at rate 
\begin{equation}\label{rate}
r(x,i|\psi)=\lambda\left\langle\psi\,|\,\Lambda_{i}(x)\psi\right\rangle
\end{equation}
and when this happens, the wave function changes to $
{\Lambda_{i} (x)^{1/2}\psi}/{\| \Lambda_{i} (x)^{1/2} \psi \|}$.

Thus, if between time $t_0$ and any time $t>t_0$, $n$ collapses have occurred at the times  $t_0< T_1 < T_2 < \ldots < T_n < t $, with centers  $X_1, \ldots, X_n$ and labels $I_1, \ldots, I_n$, the wave function at time $t$ will be
\begin{equation}\label{eq:psit}
\psi_t =  \frac{L^{F_n}_{t, t_0} \psi_{t_0}
 }{\| L^{F_n}_{t, t_0} \psi_{t_0}
\|}\,
  \end{equation}
 where $F_n = \{(X_1,T_1,I_1), \ldots, (X_n,T_n,I_n)\}$ and
\begin{equation}\label{eq:long}
L^{F_n}_{t, t_0} 
= U_{t-T_n} \Lambda_{I_n}(X_n)^{1/2} \,U_{T_n-T_{n-1}} \Lambda_{I_{n-1}}(X_{n-1})^{1/2} 
  \,U_{T_{n-1}-T_{n-2}}  \cdots \Lambda_{I_1}(X_1)^{1/2} \, U_{T_1-t_0}. 
    \end{equation}
Since $T_i$, $X_i$, $I_i$ and $n$ are random, $\psi_t$ is also random.

It should be observed that---unless $t_0$ is the initial time of the universe---also $\psi_{t_0}$ should be regarded as random, being determined by the collapses that occurred at times earlier that $t_0$. However, \emph{given} $\psi_{t_0}$, the statistics of the future evolution of the wave function is completely determined; for example, the joint distribution  of the first $n$ collapses after $t_0$, with particle labels $I_1, \ldots, I_n \in \{1,\ldots,N\}$, is
\begin{multline}\label{nflashdist}
  \PPP\bigl( X_1\in d x_1, T_1 \in d t_1, I_1 = i_1, \ldots,
 X_n \in dx_n, T_n \in d t_n, I_n = i_n   |  \psi_{t_0} \bigr) =\\ 
\lambda^n  e^{-N\lambda (t_n-t_0)} \| L^{f_n}_{t_n, t_0} \psi_{t_0}
 \|^2 \, dx_{1}dt_1   \cdots  dx_{n}dt_n \,,
\end{multline}
 with $f_n = \{(x_1,t_1,i_1), \ldots, (x_n,t_n,i_n)\}$ and $L^{f_n}_{t_n, t_0}$  given, \emph{mutatis mutandis}, by \eqref{eq:long}.

\bigskip

 This is, {more or less}, all there is to say about the formulation of the {\sf GRW} theory according to most theorists. 
In contrast, GianCarlo Ghirardi believes that the description provided above is not the whole story, and we agree with him. We believe that, depending on the choice of what we call the \emph{primitive ontology} (PO) of the theory, there are correspondingly
 different versions of the theory. We will discuss the notion of primitive
 ontology in detail in Section \ref{sec:PO}.
In the subsections below we present two versions of the {\sf GRW} theory, based on two different choices of the PO, namely the \emph{matter density ontology} (in Section~\ref{sec:GRWm}) and the \emph{flash ontology} (in Section~\ref{sec:GRWf}).

\subsection{GRWm}
\label{sec:GRWm}

In the first version of the {\sf GRW} theory, denoted by {\sf GRWm}, the PO is given by a field: We have a variable $m(x,t)$ for every point $x \in \RRR^3$ in space and every time $t$, defined by 
\begin{equation}\label{mdef}
 m(x,t) = \sum_{i=1}^N m_i \int\limits_{\RRR^{3N}}  dq_1 \cdots dq_N \, \delta(q_i-x) \,  \bigl|\psi(q_1, \ldots, q_N,t)\bigr|^2 \,.
\end{equation}
In words, one starts with the $|\psi|^2$--distribution in configuration
space $\RRR^{3N}$, then obtains the marginal distribution of 
the $i$-th degree of freedom $q_i\in \RRR^3$
by integrating out
all other variables $q_j$, $j \neq i$, multiplies by the mass associated with $q_i$, and sums over $i$.
{\sf GRWm} was essentially proposed by Ghirardi and co--workers
in \citep{Ghi};\footnote{They first proposed (for a model slightly more complicated than the one considered here)
that the matter density be given by an expression similar to \eqref{mdef} but this difference is not relevant for our purposes.} see also \citep{Gol98}.

The field $m(\cdot,t)$ is supposed to be understood as the
density of matter in space at time $t$. Since these variables are
functionals of the wave function $\psi$, they are not `hidden
variables' since, unlike the positions in {\sf BM}, they need not be specified in 
addition to the wave function, but rather are determined by it. Nonetheless, they are additional elements of the {\sf GRW} theory that need to be posited in order to have a complete description of the world in the framework of that theory. 

{\sf GRWm} is a theory about the behavior of a field $m(\cdot,t)$ on three-dimensional space. The microscopic description of reality provided by the matter density field $m(\cdot,t)$ is not particle-like but instead continuous, in contrast to the particle ontology of {\sf BM}.  
This is reminiscent of Schr\"odinger's early view of the wave function as representing a continuous matter field. But while Schr\"odinger was obliged to abandon his early view because of the tendency of the wave function to spread,  
the spontaneous wave function collapses built into the {\sf GRW} theory tend to localize the wave function, thus counteracting this tendency and overcoming the problem. 
 
A parallel with {\sf BM} begins to emerge: 
they both essentially involve more than the wave function. 
In one the matter is spread out continuously, while in the other it is
concentrated in finitely many particles; however, both theories are concerned with matter in three-dimensional space, and in some regions of space there is more than in others.

You may find {\sf GRWm} a surprising proposal. You may ask, was it not the
point of {\sf GRW} --- perhaps even its main advantage over {\sf BM} ---
that it can do without objects beyond the wave function, such as particle
trajectories or matter density? Is not the dualism present in {\sf GRWm}
unnecessary? That is, what is wrong with the version of the {\sf GRW}
theory, which we call {\sf GRW0}, which involves just the wave function and
nothing else? We will return to these questions in Section
\ref{sec:grw0}. To be sure, it seems that if there was nothing wrong with
{\sf GRW0}, then, by simplicity, it should be preferable to {\sf GRWm}. We
stress, however, that Ghirardi must regard {\sf GRW0} as seriously
deficient; otherwise he would not have proposed anything like {\sf GRWm}.
We will indicate in Section \ref{sec:grw0} why we think Ghirardi is
correct. To establish the inadequacy of {\sf GRW0} is not, however, the
main point of this paper.

\subsection{GRWf}
\label{sec:GRWf}
\begin{figure}[ht]\begin{center}\includegraphics[width=.4 \textwidth]
{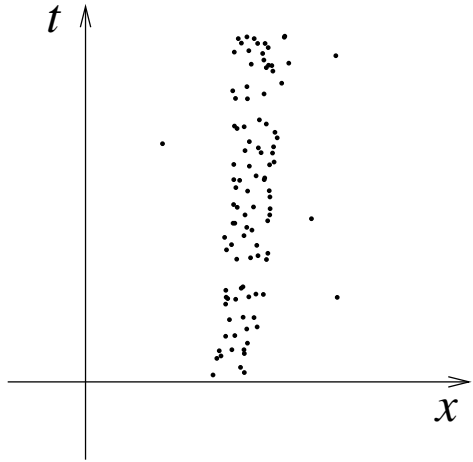}\end{center}\caption{A typical pattern of flashes in space-time, and thus a possible   
world according to the {\sf GRWf} theory}\label{flashes}\end{figure}

According to another version of the {\sf GRW} theory, which was
first suggested by \citet{Bell87,Bell89}, then adopted in \citep{kent,Gol98,Tum04,Tum05,AZ05,Mau05},
and here denoted {\sf GRWf}, the PO is given by `events' in space-time called flashes, mathematically
described by points in space-time. 
This is, admittedly, an unusual PO, but it is a possible
one nonetheless. In {\sf GRWf} matter is neither made of particles following world
lines, such as in classical or Bohmian mechanics, nor of a continuous
distribution of matter such as in {\sf GRWm}, but rather of discrete points
in space-time, in fact finitely many points in every bounded space-time
region, see Figure~\ref{flashes}. 

In the {\sf GRWf} theory, the space-time locations of the flashes can be read
off from the history of the wave function  {given by \eqref{eq:psit} and \eqref{eq:long}}: every flash corresponds to one of the spontaneous collapses of the wave function, and its space-time location is just the space-time location of that
collapse. 
Accordingly, equation \eqref{nflashdist} gives  the joint distribution  of the first $n$ flashes, after some initial   time $t_0$.
The flashes form the set
\[
  F=\{(X_{1},T_{1}), \ldots, (X_{k},T_{k}), \ldots\}
\]
(with $T_1<T_2<\ldots$).

In Bell's words:
\begin{quote}
[...] the GRW jumps (which are part of the wave function, not something else) are well localized in ordinary space. Indeed each is centered on a particular spacetime point $(x,t)$. So we can propose these events as the basis of the `local beables' of the theory. These are the mathematical counterparts in the theory to real events at definite places and times in the real world (as distinct from the many purely mathematical constructions that occur in the working out of physical theories, as distinct from things which may be real but not localized, and distinct from the `observables' of other formulations of quantum mechanics, for which we have no use here).  A piece of matter then is a galaxy of such events.
~\citep{Bell87}
\end{quote}
That is, Bell's idea is that {\sf GRW} can account for objective reality in three-dimensional space in terms of space-time points $(X_{k},T_{k})$
that correspond to the localization events (collapses) of the wave function. 
Note that if the number $N$ of the degrees of freedom in the wave function is large, as in the case of a macroscopic object, the number of flashes is also large (if $\lambda=10^{-15}$ s$^{-1}$ and $N=10^{23}$, we obtain $10^{8}$ flashes per second).
Therefore, for a reasonable choice of the parameters of the {\sf GRWf} theory, a cubic centimeter of
solid matter contains more than $10^8$ flashes per second.
That is to say that large numbers of flashes can form macroscopic shapes, such as tables and chairs. That is how we find an image of our world in {\sf GRWf}. 

Note however that at almost every time space is in fact empty, containing no flashes and thus no matter.
Thus, while the atomic theory of matter entails that space is not everywhere continuously filled with matter but rather is largely void, {\sf GRWf} entails that at most times space is entirely void.

According to this theory, the world is made of flashes and the wave function serves as the tool to generate
the `law of evolution' for the flashes: equation \eqref{rate} gives the rate of the flash process ---the probability per unit time of the flash of label $i$ occurring at the point $x$.  
 For this reason, we prefer the word `flash' to `hitting' or `collapse center': the latter words suggest that the role of these events is to affect the wave function, or that they are not more than certain facts about the wave function, whereas `flash' suggests rather something like an elementary event. Since the wave function $\psi$ evolves in a random way, $F=\{(X_k,T_k): k\in \NNN\}$ is a random subset of space-time, a point process in space-time, as probabilists would say. {\sf GRWf} is thus a theory whose output is a point process in space-time.\footnote{An anonymous referee has remarked that according to {\sf GRWf} with the original parameters, in a single living cell there might occur as few as one flash per hour, so that the cell is empty of matter for surprisingly long periods, quite against our intuition of a cell as a rather classical object. We make a few remarks to this objection. First, one should of course be careful with the language: there is presumably no cell in {\sf GRWf}, though the structure of the wave function (on configuration space---even though there are no configurations) might suggest otherwise. 
Second, it all depends on the choice of the parameters $\lambda$ and $\sigma$, and, as long as experiments have not fixed their values, this cell argument may indeed be an argument for a choice different from GRW's original one (say, with larger $\lambda$ and larger $\sigma$). We do not wish to argue here for any particular choice. 
Third, while most people might expect a cell to be real in much the same way as (say) a cat, one would not necessarily expect this of a single atom. Thus, it seems quite conceivable that, at some critical scale between that of atoms and that of cats, the ontological character of objects changes---as indeed it does in {\sf GRWf} because of the limited resolution of matter given by the space-time density of flashes (e.g., in water approximately one flash every 20 micrometers every second). The cell example shows that the critical scale in {\sf GRWf} is larger than one might have expected, and thus that {\sf GRWf} is a mildly quirky picture of the world. But this mild quirkiness should be seen in perspective. In comparison, many other views about quantum reality are heavily eccentric, as they propose that reality is radically different from what we normally think it is like: e.g., that there exist parallel worlds, or that there exists no matter at all, or that reality is contradictory in itself.}

\subsection{Empirical Equivalence Between GRWm and GRWf}
\label{sec:empequiv}

We remark that {\sf GRWm} and {\sf GRWf} are empirically equivalent, i.e., they make always and exactly the same predictions for the outcomes of experiments. In other words, there is no experiment we could possibly perform that would tell us whether we are in a {\sf GRWm} world or in a {\sf GRWf} world, assuming we are in one of the two. This should be contrasted with the fact that there are possible experiments (though we cannot perform any with the present technology) that decide whether we are in a Bohmian world or in a {\sf GRW} world.

The reason is simple. Consider any experiment, which is finished at time $t$. Consider the same realization of the wave function on the time interval $[0,t]$, but associated with different primitive ontologies in the two worlds. At time $t$, the result gets written down, encoded in the shape of the ink; more abstractly, the result gets encoded in the position of some macroscopic amount of matter. If in the {\sf GRWf} ontology, this matter is in position 1, then the flashes must be located in position 1; thus, the collapses are centered at position 1; thus, the wave function is near zero at position 2; thus, by \eqref{mdef} the density of matter is low at position 2 and high at position 1; thus, in {\sf GRWm} the matter is also in position 1,  displaying the same result as in the {\sf GRWf} world.

We will discuss empirical equivalence again in Section \ref{sec:equiv}. 

\section{Primitive Ontology }
\label{sec:PO}

The matter density field in {\sf GRWm}, the flashes in {\sf GRWf}, and the particle trajectories in {\sf BM} have something in common: they form (what we have called) the primitive ontology of these theories. The PO of a theory---and its behavior--- is what the theory is fundamentally about. 
It is closely connected with what Bell called the `local beables':
\begin{quote}
  [I]n the words of Bohr, `it is decisive to recognize that, however
  far the phenomena transcend the scope of classical physical
  explanation, the account of all evidence must be expressed in
  classical terms'. It is the ambition of the theory of local beables
  to bring these `classical terms' into the equations, and not
  relegate them entirely to the surrounding talk.~\citep{Bell76}
\end{quote}
The elements of the primitive ontology are the stuff that things are made of.
The wave function also belongs to the ontology of {\sf GRWm}, 
{\sf GRWf} and {\sf BM}, 
but not to the PO: according to these theories physical objects are not made of wave functions\footnote{We would not go so far as \citet{Fay04} and 
  \citet{stochmech1}, who have suggested that, physically, the wave function
  does not exist at all, and only the PO exists.  But we have to admit that
  this view is a possibility, in fact a more serious one than the
  widespread view that no PO exists.}. Instead, the role of the wave function in these theories is quite different, as we will see in the following.

In each of these theories, the only
reason the wave function is of any interest at all is that it is relevant to the behavior of the PO. Roughly speaking,
the wave function tells the matter how to move. In {\sf BM} the wave
function determines the motion of the particles via equation \eqref{Bohm}, in {\sf GRWm} the wave function determines the distribution of matter in the most immediate way via equation \eqref{mdef}, and in {\sf GRWf} the wave function determines the probability distribution of the future flashes via equation \eqref{nflashdist}.
 
We now see a clear parallel between {\sf BM} and the {\sf GRW}
theory, at least in its versions {\sf GRWm} and {\sf GRWf}. Each of these theories is about matter in space-time, what might be called a decoration of space-time. Each involves a dual structure $(\macroX,\psi)$: the PO $\macroX$ providing the decoration, and the wave function $\psi$ governing the \textrm{PO}. The wave function in each of these theories, which has the role of generating the dynamics for the PO, has a nomological character utterly absent in the PO. This difference is crucial for understanding the symmetry properties of these theories and therefore is vital for the construction of a Lorentz invariant quantum theory without observers, as we will discuss in Section \ref{sec:symmetry}.

Even the Copenhagen interpretation (orthodox quantum theory, {\sf OQT}) involves a dual structure: what might be regarded as its \textrm{PO} is the
classical description of macroscopic objects which Bohr insisted was
indispensable --- including in particular pointer orientations conveying the outcomes of experiments --- with the wave function serving to determine the
probability relations between the successive states of these objects. In this way, $\psi$ governs a \textrm{PO}, even for {\sf OQT}. An
important difference, however, between {\sf OQT} on the one hand
and {\sf BM}, {\sf GRWm}, and {\sf GRWf} on the other is that the latter
are fully precise about what belongs to the \textrm{PO} (particle trajectories,
respectively continuous matter density or  flashes) whereas
the Copenhagen interpretation is rather vague, even noncommittal, on this point, since the notion of `macroscopic' is an intrinsically vague one: of how many atoms need an
object consist in order to be macroscopic? And, what exactly
constitutes a `classical description' of a macroscopic object?

Therefore, as the example of the Copenhagen interpretation of quantum mechanics makes vivid, an adequate fundamental 
physical theory, one with any pretension to precision, must involve a PO defined on the microscopic scale. 

\subsection{Primitive Ontology and Physical Equivalence}
\label{sec:morePO}

To appreciate the concept of PO, it might be useful to regard the positions of particles, the mass density and the flashes, respectively, as the \emph{output} of {\sf BM}, {\sf GRWm} and {\sf GRWf}, with the wave function, in contrast, serving as part of an \emph{algorithm that generates this output}.
Suppose we want to write a computer program for simulating a system (or a universe) according to a certain theory. For
writing the program, we have to face the question: Which among the many
variables to compute should be the output of the program? 
All other variables are internal variables of the program: they may be necessary for doing the computation,
but they are not what the user is interested in. In the way we propose to understand {\sf BM}, {\sf GRWm}, and {\sf GRWf}, the output of the program, the result of the simulation, should be the particle world lines, the $m(\cdot,t)$ field, respectively the flashes; the output should look like Figure~\ref{flashes}. The wave function, in contrast, is one of the internal variables and its role is to implement the evolution for the output, the PO of the theory.

Moreover, note that there might be different ways of producing the same output, using different internal variables.
For example, two wave functions that differ by a gauge transformation generate the same law for the PO. 
In more detail, when (external) magnetic fields are incorporated into {\sf BM} by replacing all derivatives $\nabla_k$ in \eqref{Bohm} and \eqref{Schr} by $\nabla_k - i e_kA(q_k)$, where $A$ is the vector potential and $e_k$ is the electric charge of particle $k$, then the gauge transformation
\begin{equation}
  \psi \mapsto e^{i\sum_k e_kf(q_k)} \psi\,, \quad A \mapsto A +\nabla f
\end{equation}
does not change the trajectories nor the quantum equilibrium
distribution.
As another example, one can write the law for the PO in either the Schr\"odinger or the Heisenberg picture. As a consequence, the same law for the PO is generated by either an evolving wave function and static operators or a static wave function and evolving operators.
In more detail, {\sf BM} can be reformulated in the Heisenberg
picture by rewriting the law of motion as follows:
\begin{equation} 
  \frac{dQ_i}{dt} = - \frac{1}{\hbar} \Im \frac{\sp{\psi}{P(dq,t)
  [H,\widehat{Q}_i(t)]|\psi}}{\sp{\psi}{P(dq,t)|\psi}} (q=Q(t))\,,
\end{equation}
where $H$ is the Hamiltonian (e.g., for $N$ spinless particles given by \eqref{eq:H}), $\widehat{Q}_i(t)$ is the (Heisenberg-evolved) position operator (or, more precisely, triple of operators corresponding to the three dimensions of physical space) for particle $i$ and $P(\cdot,t)$ is the projection--valued measure (PVM) defined by the joint spectral decomposition of all (Heisenberg-evolved) position operators \citep{crea2B}.

We suggest that two theories be regarded as \emph{physically equivalent} when they lead to the same history of the PO. Conversely,
one could define the notion of PO in terms of physical equivalence: The PO is described by those variables which remain invariant under all physical equivalences.
We discuss this issue in more detail in Sections \ref{sec:lgrwf} and \ref{sec:BMC}, when presenting some examples.

What is `primitive' about the primitive ontology? That becomes clear when we realize in what way the other objects in the theory (such as the wave function, or the magnetic field in classical physics) are non--primitive: One can explain what they are by explaining how they govern the behavior of the PO, while it is the entities of the PO that make direct contact with the world of our experience. We can directly compare the motion of matter in our world with the motion of matter in the theory, at least on the macroscopic scale. The other objects in the theory can be compared to our world only indirectly, by the way they affect the PO.\footnote{While the notion of PO is similar to Bell's notion of local
beables, it should be observed that not all local beables, such as the
electric and magnetic fields in classical electrodynamics, need to be regarded as part
of the PO. Moreover, the very
conception that the PO must involve only local beables (i.e., be represented by 
mathematical objects grounded in familiar three-dimensional space)
could in principle be questioned; this is, however, a rather delicate and
difficult question that will be briefly addressed in Section \ref{sec:PO3N}
 but that deserves a thorough analysis that will be undertaken
in a separate work \citep{semicolon}.} 
 
\subsection{Primitive Ontology and Symmetry}
\label{sec:symmetry}
{The} peculiar flash ontology was invented by Bell in 1987 as a step
toward a relativistic {\sf GRW} theory. He wrote in \citep{Bell87} about {\sf GRWf}:
\begin{quote}
I am particularly struck by the fact that the model is as Lorentz
  invariant as it could be in the nonrelativistic version. It takes
  away the ground of my fear that any exact formulation of quantum
  mechanics must conflict with fundamental Lorentz invariance.
\end{quote}
What Bell refers to in the above quotation is the following. 
An analogue of the relativity of simultaneity, i.e. of the invariance of the dynamics under boosts, in the framework of a nonrelativistic theory is the invariance under relative time translations for two very distant systems. \citet{Bell87,Bell89} verified by direct calculation that {\sf GRWf} has this symmetry. However, it
it is important here to appreciate
what this invariance means. 
To say that a theory has a given symmetry is to say that 
\begin{quote}
\emph{The possible histories of the PO, those that are allowed by the theory, when transformed according to the symmetry, will again be possible histories for the theory, and the possible probability distributions on the histories, those that are allowed by the theory,  when transformed according to the symmetry, will again be  possible probability distributions for the theory}.
\end{quote}
Let us explain.
\begin{itemize}
\item `{\em The possible histories of the PO, those that are allowed by the theory\ldots}'
{We give some examples, involving Galilean 
invariance.} In classical mechanics the meaning {is} straightforward: the PO is that of particles, described by their positions in  physical space, {a history of this PO corresponds to a collection of particle trajectories---the trajectories  $Q_i(t)$, $i=1, \ldots ,N$, in 
a universe  of $N$ particles---and a history
is allowed if the particles obey Newton's law, i.e., if} $m_i \ddot Q_i(t) = F_i (Q_1(t), ..., Q_N(t))$, where $F_i$ is the Newtonian force acting on the $i$th particle. The theory is defined once the
form of  $F_i$ is specified (for example, that the force is the Newtonian gravitational force).

Consider now {\sf BM}: also here the PO is that of particles  and a possible history of the PO---one that is allowed by {\sf BM}---is a history described by  the particle trajectories 
$Q_i(t)$, $i=1, \ldots N$, which satisfy equation \eqref{Bohm} {\em for some wave function $\psi$}  satisfying  equation \eqref{Schr}. The theory is defined once the Hamiltonian $H$ in  \eqref{Schr} is specified (for example, as given by \eqref{eq:H}, for a choice of the potential $V$).

\item `{\em \ldots when transformed according to the symmetry\ldots}' 
Since the PO is represented by
a geometrical entity in physical space (a decoration of space-time, as we have said earlier), space-time symmetries naturally act on it, for example 
transforming trajectories $Q_i(t)$ to trajectories $\tilde{Q}_i(t)$. {For example,} 
under a Galilean boost (by a relative velocity $v$), in classical mechanics as well as in {\sf BM}, the trajectories $Q_i(t)$ transform into the boosted trajectories $\tilde{Q}_i(t) = Q_i(t) + v t$.

\item  `{\em \ldots  will again be possible histories for the theory\ldots}'
Notice that $Q_i(t)$ and $\tilde{Q}_i(t)$ may {arise in {\sf BM} from} different wave functions. In other words, the wave function must also be transformed when transforming the history of the PO.   However, while there is a natural transformation of  the history of the PO, there  is {not necessarily a} corresponding natural change of   the wave function.    The latter is  allowed to change  in any  way, solely determined by its relationship to the PO. {For example, consider} again a Galilean boost (by a relative velocity $v$) in {\sf BM}:
the boosted trajectories $\tilde{Q}_i(t) = Q_i(t) + v t$ form again a solution of \eqref{Bohm} and \eqref{Schr} with $\psi$ replaced by the transformed wave function\footnote{
Under this transformation $V=V(q_1, \ldots, q_N)$ in \eqref{Schr} must be replaced by $\tilde{V}=V(q_1-vt, \ldots, q_N-vt)$. For 
$V$ arising from the standard two-body interactions, {we have that $V=\tilde{V}$, and hence} the theory is invariant.}
\begin{equation}\label{Galilean}
  \tilde{\psi}_t(q_1, \ldots, q_N) = \exp\Bigl(\tfrac{i}{\hbar}
  \sum_{i=1}^N m_i
  (q_i \cdot v - \tfrac{1}{2} v^2 t)\Bigr) \, \psi_t(q_1-v t,
  \ldots, q_N-v t).
\end{equation}
Since the trajectories of the PO transformed according to the symmetry are still solutions, {\sf BM} is symmetric under Galilean transformation, even though the corresponding wave function has to undergo more than a simple change of variables in order to make this possible. 

\item `{\em \ldots and the possible probability distributions on the histories, those that are allowed by the theory\ldots}'
In a deterministic theory, {a probability distribution on the histories arises from a probability distribution on the initial conditions. In {\sf BM}, a probability distribution on histories is possible 
if there exists a wave function $\psi$ such that
the given distribution is the one induced on solutions to \eqref{Bohm} by the probability distribution $|\psi(q_1, \ldots, q_N)|^2$ at some initial time.}

More interesting is the case of nondeterministic theories. For these theories, i.e., for theories  involving stochasticity  at the fundamental level,  the law for the PO {amounts} to a specification of possible probability distributions, for example {by specifying the generator, or transition probabilities, of a Markov process.} 
For example, in  {\sf GRWm} the {history of the} PO is the mass density {field $m(\cdot, \cdot)$, and a probability distribution on the histories of this PO is possible if it is the distribution induced on $m(\cdot, \cdot)$,} according to equation \eqref{mdef},  {by some wave function $\psi$}  with probability law given, say, by \eqref{nflashdist} (and \eqref{eq:psit}). 
The case of {\sf GRWf} is analogous: {a probability distribution for the flashes 
$F=\{(X_k,T_k):k \in \NNN\}$ is possible if} induced by \eqref{nflashdist} for some wave function $\psi$.

\item `{\em\ldots  when transformed
according to the symmetry, will again be possible probability distributions for the theory.}' 
The probability distribution on the histories, when transformed according to the symmetry, is the distribution of the transformed histories. {In other words, the action of a transformation on every history determines the transformation of a probability distribution on the space of histories. As in the deterministic case,} the wave function is allowed to change in any way compatible with its relationship to the {PO}.   
For example, consider the Galilean invariance of {\sf GRWf}: 
let  $\psi$ and $\tilde{\psi}$ be two initial wave functions related as in \eqref{Galilean}, that is, by the usual formula for Galilean transformations in quantum mechanics. Let $G_t$ denote the transformation operator {in \eqref{Galilean}} at time $t$, such that $\tilde{\psi}_t = G_t \psi_t$. A simple calculation shows that
$$
\Lambda_i(x+vt)^{1/2} G_t = G_t \Lambda_i(x)^{1/2}.
$$
As a consequence, the distribution (7) of the {(spatial location of the)} first flash {arising} from $\tilde{\psi}_{T_1}$  is that {arising} from $\psi_{T_1}$ shifted by $vT_1$, and the post-collapse wave functions (6) are
still related by the appropriate $G_t$ operator, i.e.,
$$
\tilde{\psi}_{T_1+} = G_{T_1} \psi_{T_1+}.
$$
Thus, the joint distribution of flashes arising from $\tilde{\psi}$ is just the one arising from $\psi$ shifted by $vt$ for every $t$.
\end{itemize}

Going back to the work of Bell {mentioned in the beginning of this section \citep{Bell87}, what} Bell had to do for {\sf GRWf}, and what he did, was to confirm the invariance under relative time translations of the stochastic law for $F=\{(X_k,T_k):k \in \NNN\}$, the galaxy of flashes. And more generally the invariance of {\sf GRWf} directly concerns the stochastic law for the PO; 
it concerns the invariance of the law for the wave function only indirectly, contrary to what is  often, 
erroneously, believed. Under a space-time symmetry the PO must be transformed in accord with its intrinsic geometrical nature, while wave functions (and other elements of the non-primitive ontology, if any) should be transformed in a manner dictated by their relationship to the PO.

Moreover, note that there is no reason to believe that when changing the PO of a theory the symmetry properties of the theory will remain unchanged. Actually, the opposite is true.
This fact was pointed out in \citep{Gol98} and has recently been emphasized also in \citep{Tum04}, in which it has been shown that {\sf GRWf}, without interaction, can  
 be modified so as to become a relativistic quantum theory without 
 observers.\footnote{To put this result into perspective, note that the absence of interaction does not make the problem trivial. On the contrary, the main difficulty with devising a relativistic quantum theory without observers arises already in the non-interacting case: To specify a law for the PO that is relativistic but nonlocal. Note further that it would not have sufficed to specify a Lorentz-invariant evolution law for $\psi$ (entailing suitable collapse) while leaving open the law for the PO. Finally, note that for {\sf GRWm} and {\sf BM} it is not known how to 
make them ``seriously'' relativistic, i.e. without the incorporation of
additional structure that yields a foliation of space-time.} In that paper the stochastic law for the galaxy of the flashes in space-time, the PO of {\sf GRWf}, with suitably modified, Lorentz-invariant equations, has been shown explicitly to be relativistically invariant \citep[see also][]{Tum06}. 
Hence, {\sf GRWf} is Lorentz invariant, but {\sf GRWm} is not. Thus, one should not
ask whether {\sf GRW} as such is Lorentz invariant, since the answer to this question depends on
the choice of PO for {\sf GRW}. For details see \citep{Mau05}. 
Similar results to those of \citep{Tum04} have been obtained also by \citet{Fay02}
for a relativistic collapse theory on the lattice \citep[see also][]{Fay03,Fay04}.

We conclude with some remarks on {\sf OQT}.  Here the relevant PO {consists, rather vaguely of course,} of the `pointer variables' registering the results of experiments that are spoken of as measurements of quantum observables. Though {\sf OQT} provides neither {detailed histories of the PO} nor probability distributions thereof,
it does provide probability distributions  for the results of measurements registered by the 
PO, which are  given by the appropriate spectral measures for the self-adjoint operators representing the observables. In particular, the mean value of the result of the measurement $\mathcal{E}$ of the quantum observable represented by the self-adjoint operator $A$ for a system 
in the state $\psi$ is
\begin{equation} <A>_{\psi} =  \frac{\langle \psi\,| A\psi\rangle}{\langle\psi\,|\,\psi\rangle}\,.
\label{meanvs}
\end{equation}

Now consider the action of a symmetry on the experiment ${\mathcal{E}}$: it transforms $\mathcal{E}$ to the experiment $\mathcal{\tilde{E}}$ arising from the natural action of the symmetry on the physical processes defining $\mathcal E$. If $\mathcal{E}$ is a measurement of the operator $A$---that is, if the probability distribution of the results of $\mathcal{E}$ are given by the spectral measures for $A$---then $\mathcal{\tilde{E}}$ will be a measurement of the operator $\tilde{A}$ arising from $A$ under the symmetry. While $\mathcal E$ and $\mathcal{\tilde{E}}$ are of course (usually) different experiments, it is obvious from their relationship that the distribution of the results of $\mathcal{E}$ when the system is in the state $\psi$ is the same as the distribution of the results of $\mathcal{\tilde{E}}$ when the transformed system is in the transformed state $\tilde{\psi}$. In particular, where $\mathcal{E}$ is a measurement of $A$, we have that 
\begin{equation} <A>_{\psi}=  \frac{\langle \psi\,| A\psi\rangle}{\langle\psi\,|\,\psi\rangle}
=   \frac{\langle \tilde{\psi}\,| \tilde{A}\tilde{\psi}\rangle}{\langle\tilde{\psi}\,|\,\tilde{\psi}\rangle} =<\tilde{A}>_{\tilde{\psi}} \, .
\label{meanv2}
\end{equation}
According to the analysis of \citet{wigner} and \citet{bargmann}, these  transformations  on wave functions and operators are given by  unitary or anti-unitary 
operators $U$, i.e.,  $\tilde{\psi}= U\psi$, $\tilde{A}= UAU^{-1}$,  where $U$ is an element of a unitary-projective representation of the symmetry group.

Note that while the distribution of the {\em result} of the experiment is, for trivial reasons, unaffected by the symmetry transformation, the macroscopic PO is in fact transformed. For example, a rotated experiment will involve a rotated `pointer position,' or a rotated computer printout. But what the pointer is pointing to, and what the printout says, will not change. 
In other words, the numerical result $Z$ of an experiment $\mathcal{E}$
should not be confused  with the macroscopic  configuration $M$ of the
pointer variables, the PO of {\sf OQT}, the former being indeed a function of the
latter,
i.e., $Z=f(M)$, with the function $f$ expressing  the `calibration' of the
experiment. In  $\mathcal{\tilde{E}}$, the rotated experiment, the
PO (the pointer orientation)  changes together with the calibration: the
pointer points in a different direction $\tilde{M}$ and the scale $f$  is
rotated into $\tilde{f}$ such that  $\tilde{f}(\tilde{M})=f(M)$.

Thus, when all is said and done, although the PO of {\sf OQT} is rather vague and imprecise, insofar as symmetry is concerned the situation is  indeed  analogous to that of theories, such as {\sf BM} or {\sf GRWf},  having  a clear and exactly specified  PO: also for {\sf
 OQT} the possible probability distributions on the PO, those that are allowed by the theory,  when transformed according to the symmetry, will again be  possible probability distributions for the theory.

\subsection{Without Primitive Ontology}
\label{sec:grw0}

Now let us turn to the question: What is wrong with {\sf GRW0}, the bare version of {\sf GRW}, which involves just the wave function and nothing else? Why does one need a PO at all? 
Our answer is that we do not see how the existence and behavior of
 tables and chairs and the like could be accounted for without positing a
 primitive ontology---a description of matter in space and time.

 The aim of a fundamental physical theory is, we believe, to describe the
 world around us, and in so doing to explain our experiences to the extent
 of providing an account of their macroscopic counterparts, an account of
 the behavior of objects in $3$-space. Thus it seems that for a fundamental
 physical theory to be satisfactory, it must involve, and fundamentally be
 about, `local beables,' and not just a beable such as the wave function,
 which is non-local: In the words of \citet{Bell87}
\begin{quote}
  [...] the wave function as a whole lives in a much bigger space, of 
  $3N$ dimensions. It makes no sense to ask for the amplitude or 
  phase or whatever of the wave function at a point in ordinary space. 
  It has neither amplitude nor phase nor anything else until a 
  multitude of points in ordinary three-space are specified.
\end{quote}

In contrast, if a law is, like the {\sf GRW} process in Hilbert space,
about a mathematical object, like $\psi$, living in some abstract space,
like $\RRR^{3N}$, it seems necessary to have or to add something more in
order to make contact with a description in 3-space. For example,
formulations of classical mechanics utilizing configuration space
$\RRR^{3N}$ or phase space $\RRR^{6N}$ (such as Euler--Lagrange's or
Hamilton's) are connected to a PO in 3-space (particles with trajectories)
by the definitions of configuration space and phase space.

If, as we believe, a PO given by local beables is so crucial for a theory
to make sense as a fundamental physical theory, one might wonder how {\sf
  GRW0} could be taken seriously by so many serious people \citep[see,
e.g.,][]{albert,albert1,rimini,plewis}. One reason, perhaps, is that if the
wave function $\psi$ is suitably collapsed, i.e., concentrated on a subset
$S$ of configuration space such that all configurations in $S$ look
macroscopically the same, all corresponding for example to a pointer
pointing in the same way, then we can easily imagine what a world in the
state $\psi$ is macroscopically like: namely like the macrostate defined by
configurations from $S$. For example, when in {\sf{GRW0}} the wave function
is concentrated near $q$, where $q$ is a configuration describing a pointer
pointing to the value $a$, it is easy to feel justified in concluding that
there is a pointer that is pointing to the value $a$, and to forget that we
are dealing with a theory for which there exists no arrangement of stuff in
physical three-dimensional space at all. 

Since the macroscopic description
does not depend on whether the PO consists of world lines, flashes, or a
continuous distribution of matter, and since the reasoning does not even
mention the PO, it is easy to overlook the fact that a further law needs to
be invoked, one which prescribes how the wave function is related to the
PO, and implies that for wave functions such as described, the \textrm{PO}
is such that its macroscopic appearance coincides (very probably) with the
macroscopic appearance of configurations in $S$. To overlook this step is
even easier when focusing very much on the measurement problem, whose
central difficulty is that the wave function of object plus apparatus, if
it evolves linearly, typically becomes a superposition of macroscopically
distinct wave functions like $\psi$ which thus contains no hint of the
actual outcome of the experiment.

\label{sec:PO3N}

Interestingly enough, after having underlined the importance of local beables for a fundamental physical theory, Bell proposed {\sf GRW} to be about `stuff' in configuration ($3N$-dimensional)
space. In his celebrated analysis of the quantum measurement problem \citep{Bell90}, he wrote:
\begin{quote}
  The GRW-type theories have nothing in their kinematics but the
  wavefunction. It gives the density (in a multidimensional
  configuration space!) of \emph{stuff}. To account for the narrowness
  of that stuff in macroscopic dimensions, the linear Schr\"odinger
  equation has to be modified, in the GRW picture by a mathematically
  prescribed spontaneous collapse mechanism. [Emphasis in the
  original.]
\end{quote}
He made a similar remark to Ghirardi (quoted by the latter in \citep[p.~345]{BG03}) in a letter dated October 3, 1989: 
\begin{quote}
  As regards $\psi$ and the density of stuff, I think it is important that this density is in the $3N$-dimensional configuration space. So I have not thought of relating it to ordinary matter or charge density in $3$-space.  Even for one particle I think one would have problems with the latter. So I am inclined to the view you mention `as it is sufficient for an objective interpretation' ... And it has to be stressed that the `stuff' is in $3N$-space---or whatever corresponds in field theory. 
\end{quote} 

As we have indicated already, we don't understand this proposal, which
clearly suffers from the difficulties discussed above.  Whoever suggests
that matter exists not in $3$-space but in $3N$-space must bridge the gap
between an ontology in $3N$-space and the behavior of objects in $3$-space.
Strategies for doing so have in fact been proposed; see \citep{albert1} for
a proposal and
\citep{monton} for a critique. For the reasons mentioned above, we do not
believe that they can succeed.

\subsection{Primitive Ontology and Quantum State}
It is well known that in {\sf OQT} the quantum state is naturally projective. {That is,} quantum states are best regarded as mathematically represented by {\em rays} in the system's Hilbert space $\mathscr{H}$, i.e. by the elements  of the projective space $\mathbb{P}(\mathscr{H})$, consisting of equivalence classes of wave functions $\psi\in \mathscr{H}$ differing by a multiplicative constant. 
{This follows from} the rule \eqref{meanvs}
for the mean value of an observable represented by a self-adjoint operator $A$ for a system 
in the state $\psi$. 
Wave functions $\psi$  differing by a multiplicative constant give the same mean value to all observables $A$.

Similarly, in {\sf BM} the quantum state is naturally projective: it follows from \eqref{Bohm} that wave functions differing by a
   multiplicative constant are associated with the same vector field, and thus generate the same dynamics for the PO.\footnote{And insofar as probabilities are concerned, if $\psi$ is not normalized, these are given by $|\psi(q)|^2/\langle\psi|\psi\rangle$, which is projective.} 
 
In {\sf GRWf} the quantum state is also naturally projective. Of course, for general $\psi$ (not necessarily normalized), instead of  \eqref{rate} the rate for the flashes should be given by
   \begin{equation}\label{raten}
r(x,i|\psi)=\lambda \frac{\left\langle\psi\,|\,\Lambda_{i}(x)\psi\right\rangle}{ \left\langle\psi\,|\psi\right\rangle}\,.
\end{equation}

In {\sf GRWm} wave functions differing by a multiplicative constant of modulus 1 define the same evolution of the mass density field  \eqref{mdef}. If the wave function is multiplied by a more general constant,  in order to ensure the same evolution of the mass density
the right hand side of \eqref{mdef} could be divided by $\langle\psi|\psi\rangle$. But this is perhaps unnecessary, since universal mass densities that differ only by a multiplicative constant are arguably physically equivalent.

{\sf GRW0}, involving only wave functions, does not allow us to make the 
same kind of argument; it is thus not clear for {\sf GRW0} why $\psi$ should be regarded 
as projective, though the structure of {\sf GRW0} is compatible with doing 
so.

To sum up, the projective nature of the quantum state can be regarded as a consequence of the axioms of {\sf OQT}, {\sf BM}, {\sf GRWm} and {\sf GRWf}, but not of {\sf GRW0}.

\section{Differences between BM and GRW}
\label{sec:diff}

We have stressed the similarity between {\sf BM} and {\sf GRW}.  There are, of course, also significant differences. Perhaps the most obvious is that in {\sf BM} the Schr\"odin\-ger evolution is exact, but not in {\sf GRW}. However, this difference is not so crucial. In fact we will present in Section \ref{sec:lgrwf} a reformulation of {\sf GRWf} in which the Schr\"odin\-ger evolution is exact.

A related important difference is that the empirical
predictions of {\sf BM} agree \emph{exactly and always} with
those of the quantum formalism (whenever the latter is unambiguous) while the predictions of the {\sf GRW} theory don't. (The latter agree only approximately and in most cases.)
In particular, one can empirically distinguish {\sf BM} from the {\sf GRW}
theory. (However, no decisive test could as yet be performed; see
\citep{BG03} for details.)
The empirical disagreement between the two theories is usually explained by appealing to the fact that in one theory the wave function obeys the Schr\"odin\-ger evolution while in the other it does not. However, especially in light of the reformulation of {\sf GRWf} we shall describe  in Section \ref{sec:lgrwf}, the empirical inequivalence between the two theories should be better regarded as having a different origin. Though we shall elaborate on this issue in  Section \ref{sec:equiv},  we shall anticipate  the mathematical roots of such a difference in Section  \ref{subsec:poae} (which however may be skipped on a first reading of this paper).

A difference in the mathematical structure of {\sf GRWf} (and {\sf OQT}) on the one hand and {\sf BM} (but also {\sf GRWm}) on the other concerns the probability distribution 
that each of the these theories defines on its space of histories of the PO. 
This probability distribution is a quadratic functional of the initial $\psi$ for {\sf GRWf} and {\sf OQT}, but not for {\sf BM} and {\sf GRWm}. This feature is at the origin of why {\sf GRWf} can be modified so as to become a fully relativistically invariant theory (see the end of Section \ref{sec:symmetry}). It will be discussed in the following subsection, which, however, will not be needed for understanding the rest of the paper.

\subsection{Primitive Ontology and Quadratic Functionals}
\label{sec:appendix}
It is worth noting a feature of the mathematical structure of {\sf GRWf} that it shares with {\sf OQT}, but that is absent in, for example, {\sf BM} and {\sf GRWm}. It concerns the dependence on the (initial) wave function $\psi$ of the probability distribution $\PPP^{\psi}$ that the theory defines on its space $\Omega$ of histories of the PO. 
In {\sf BM}, $\Omega$ is the space of continuous paths in configuration space $\RRR^{3N}$, and the measure $\PPP^\psi$ corresponds to the quantum equilibrium measure, and is concentrated on a
$3N$-dimensional submanifold of $\Omega$, namely the solutions of Bohm's equation \eqref{Bohm}. In {\sf GRWf}, $\Omega$ is the space of discrete subsets of space-time (possibly with labels $1,\ldots,N$), and the measure $\PPP^\psi$ is given by \eqref{nflashdist}. In {\sf GRWm}, $\Omega$ is a space of fields on space-time, and $\PPP^\psi$ the image under the mapping $\psi \mapsto m$ given by \eqref{mdef} of the distribution of the Markov process $(\psi_t)_{t \geq 0}$.

In {\sf GRWf} and {\sf OQT}, but not in {\sf BM} or {\sf GRWm}, $\PPP^\psi$ is a quadratic functional of $\psi$. More precisely, in {\sf GRWf} and {\sf OQT} it is of the form
\begin{equation}\label{PPOVM}
  \PPP^\psi(\cdot) = \sp{\psi}{E(\cdot) \psi}
\end{equation}
where $E(\cdot)$ is the positive-operator-valued measure (POVM) on $\Omega$ that can be read off from \eqref{nflashdist} for {\sf GRWf}, and is the POVM associated with the results of a sequence of measurements for {\sf OQT}  \citep[see, e.g.,][]{DGZ04}.
Neither {\sf GRWm} nor {\sf BM} share this property. The easiest way of seeing this begins with noting that \eqref{PPOVM} entails that any two ensembles of wave functions (corresponding to probability measures $\mu,\mu'$ on the unit sphere $S$ of Hilbert space) with the same density matrix,
\begin{equation}\label{dm}
  \hat{\rho}_\mu = \int_S \mu(d\psi) \, |\psi \rangle \langle \psi| =  \hat{\rho}_{\mu'} \,,
\end{equation}
lead to the same distribution
\begin{equation}\label{sameP}
  \PPP_\mu(\cdot) = \int_S \mu(d\psi) \, \PPP^\psi(\cdot) =  \mathrm{tr}(E(\cdot) \hat{\rho}_\mu) = \PPP_{\mu'}(\cdot)
\end{equation}
on $\Omega$. This is notoriously not true in {\sf BM} \citep{Bell80}. It is not true in {\sf GRWm} either, as one easily checks, for example by considering, at just one single time, the following two ensembles of wave functions for Schr\"odinger's cat: $\mu$ gives probability $\frac{1}{2}$ to $2^{-1/2} (|\mathrm{dead}\rangle + |\mathrm{alive}\rangle)$ and $\frac{1}{2}$ to $2^{-1/2} (|\mathrm{dead}\rangle - |\mathrm{alive}\rangle)$, while $\mu'$ gives $\frac{1}{2}$ to $|\mathrm{dead}\rangle$ and $\frac{1}{2}$ to $|\mathrm{alive}\rangle$.

One can say that the essence of this difference between these theories lies in different choices of which quantity is given by a simple, namely quadratic, expression in $\psi$:
\begin{itemize}
\item the probability distribution $\PPP^\psi$ of the history of the PO both in {\sf GRWf} and {\sf OQT}, see \eqref{PPOVM}
\item the probability distribution $\rho^\psi$ of the PO at time $t$ in {\sf BM},
\begin{equation}
  \rho^\psi(q,t) = |\psi(q,t)|^2
\end{equation}
\item the PO itself at time $t$ in {\sf GRWm},
\begin{equation}
  m(x,t) = \sp{\psi_t}{\tilde{\Lambda}(x) \, \psi_t} \text{ with }
  \tilde{\Lambda}(x) = \sum_{i=1}^N m_i \, \delta(x-\widehat{Q}_i)\,.
\end{equation}
\end{itemize}
Note in particular the rather different roles that `$|\psi|^2$' can play for different quantum theories and different choices of the PO. 

\subsection{Primitive Ontology and Equivariance}
\label{subsec:poae}

In Section \ref{sec:bm} we have recalled the notion of the equivariance of
the probability distribution $|\psi|^2$ and indicated how it is the key
notion for establishing the empirical agreement between {\sf BM} and the
predictions of the quantum formalism (whenever the latter are unambiguous).
The equivariance of $|\psi|^2$ expresses the mutual compatibility,
with respect to $|\psi|^2$, of the Schr\"odinger
evolution of the wave function and the Bohmian motion of the configuration.

It would seem natural to expect that for {\sf GRWf} we also have
equivariance, but relative to the (stochastic) {\sf GRW} evolution of the
wave function instead of the Schr\"odinger evolution. However, the concept
of the equivariance of the distribution $|\psi|^2$ is not directly
applicable in this case: in fact, for {\sf GRWf} there is no random
variable $Q(t)$ whose distribution could agree or disagree with a
$|\psi_t|^2$ distribution (or any other quantum mechanical distribution),
since {\sf GRWf} is a theory of flashes, not particles, and as such yields
no nontrivial random variable that can be regarded as associated with a
fixed time $t$.  In this framework it seems natural to consider the notion
of a time--translation equivariant distribution, in terms of which we may
provide a generalized notion of equivariance as follows: Let $\Omega_t$ be
the space of possible histories of the PO for times greater than or equal
to $t$. In trajectory theories like {\sf BM}, $\Omega_t$ is the space of
continuous paths $[t,\infty) \to \Q$, where $\Q$ is the configuration
space; in flash theories like {\sf GRWf} it is the space of
finite--or--countable subsets of the half space-time $[t,\infty) \times
\RRR^3$.
Consider an association $\psi \mapsto \PPP^{\psi}$ where
$\PPP^{\psi}$ is a probability measure on $\Omega_0$ that is compatible
with the dynamics of the theory.  We say that this association is
\textit{equivariant} relative to a deterministic evolution $\psi \mapsto
\psi_t$ if $S_t^{\star} \PPP^{\psi} = \PPP^{\psi_t}$, where ${\star}$
denotes the action of the mapping on measures and $S_t$ is a suitably defined time shift.\footnote{In order to define $S_t$ properly, let $R_t,\ t>0$, be the restriction mapping $\Omega_0 \to
\Omega_t$, and $T_{\tau}$ the time translation mapping $\Omega_t \to
\Omega_{t+\tau}$. Then $S_t = T_{-t} \circ R_t : \Omega_0 \to \Omega_0$ is
the time shift.}
More generally, for an
evolution that may be stochastic, we say that the association is
\textit{equivariant} relative to the evolution if
\begin{equation}S_t^{\star} \PPP^{\psi} = \EEE \PPP^{\psi_t},\end{equation}
 where $\EEE$ denotes the average over the random $\psi_t$. 
With this definition, {\sf BM} is equivariant relative to the Schr\"odinger evolution, and {\sf GRWf} and {\sf GRWm} are equivariant relative to the {\sf GRW} evolution.

\section{A Plethora of Theories}
\label{sec:differences}
One may  wonder whether some primitive ontologies (flashes and
continuous matter density) work only with {\sf GRW}-type theories while others
(particle trajectories) work only with Bohm-type theories. This is  
not the case, as we shall explain in  this section.

\subsection{Particles, Fields, and Flashes}

Let us analyze, with the aid of Table 1, several possibilities:
\begin{table}[t!]
\begin{center}
\begin{tabular}{|c||c|c|c|}\hline  & Particles & Fields & Flashes \\\hline\hline Deterministic & {\sf BM} & {\sf BQFT}, {\sf Sm} &  \\\hline Indeterministic & {\sf SM}, {\sf BTQFT}, {\sf BMW}, {\sf GRWp} & 
{\sf GRWm} &  {\sf GRWf}, {\sf Sf}, {\sf Sf$'$} \\\hline \end{tabular} \caption{\small{Different possibilities for the PO of a theory are presented: particles, fields and flashes. These different primitive ontologies can evolve according to either deterministic or stochastic laws.  
Corresponding to these possibilities we have a variety of physical theories: Bohmian mechanics ({\sf BM}), Bohmian quantum field theory ({\sf BQFT}),   a mass density field theory with Schr\"odinger evolving wave function ({\sf Sm}), stochastic mechanics ({\sf SM}), Bell-type quantum field theory ({\sf BTQFT}), Bell's version of many-worlds ({\sf BMW}), a particle {\sf GRW} theory ({\sf GRWp}),
{\sf GRW} theory with mass density ({\sf GRWm}), {\sf GRW} theory with flashes ({\sf GRWf}), 
and two theories with flashes governed by Schr\"odinger (or Dirac) wave functions ({\sf Sf} and {\sf Sf$'$}).  For a detailed description of these theories, see the text.}}
\end{center}
\label{defaulttable}
\end{table}
there can be at least three different kinds of primitive ontologies for a fundamental physical theory, namely particles, fields, and flashes. Those primitive ontologies can evolve either according to a deterministic or to a stochastic law and this law can be implemented with the aid of a wave function evolving either stochastically or deterministically.

{\sf BM} is the prototype of a theory in which we have a particle ontology that evolves deterministically according to a law specified by a wave function that also evolves deterministically. The natural analog for a theory with particle ontology with indeterministic evolution is stochastic mechanics ({\sf SM}), in which the law of evolution of the particles is given by a diffusion process while the evolution of the wave function, the usual Schr\"odin\-ger evolution, remains deterministic 
\citep[see][for details]{stochmech1,stochmech2}.
Another example involving stochastically evolving particles with a deterministically evolving wave function is provided by a Bell-type quantum field theory ({\sf BTQFT}) in which, despite the name, the PO is given by particles evolving indeterministically to allow for creation and annihilation \citep[for a description, see][]{crlet,crea2B,Bell86}. 
Another possibility for a stochastic theory of particles is a theory {\sf GRWp} in which the particle motion is governed by \eqref{Bohm} but with a wave function that obeys a {\sf GRW}-like  evolution in which the collapses occur exactly as in {\sf GRW} except that, once the time and label for the collapse has been chosen, the collapse is centered at the actual position of the particle with the chosen label, rather than at random according to equation \eqref{p}. (A garbled formulation of this theory is presented in \citep[p. 346]{BohmHiley}.)

What in Table 1 we call a Bohmian quantum field theory ({\sf BQFT}) involves only fields, evolving deterministically \citep{Bohm52,westman}.
Another example is provided by the theory {\sf Sm} in which the PO is given by the mass density field \eqref{mdef} but evolving with a Schr\"odinger wave function --- always evolving according to Schr\"odinger's equation, with no collapses.   
{\sf GRWm} provides an example of a theory of fields that evolve stochastically. 

Concerning theories with flashes, these are inevitably stochastic, and {\sf GRWf}, in which the flashes track the collapses of the wave function, is the prototype. However, there are also theories with flashes in which the wave function never collapses. Such theories are thus arguably closer
to {\sf BM} than to {\sf GRWf}. We consider two examples.

In the first example, denoted by {\sf Sf},\footnote{Here {\sf S} stands for Schr\"odinger (evolution). Using this notation we have that {\sf BM} = {\sf Sp}.} the PO consists of flashes with their distribution determined by a Schr\"odinger wave function $\psi =\psi (q_1, \ldots, q_N)$, that evolves always unitarily, as in {\sf BM}, according to the $N$--`particle' Schr\"odinger evolution \eqref{Schr}.
The flashes are generated by the wave function exactly as in {\sf GRWf}. Thus, the algorithm, whose output is the flashes, is the same as the one described in Section \ref{sec:GRW}, with steps 1., 2. and 3., with the following difference: the first sentence in step 2. is dropped, since no collapse takes place. 
In other words, in {\sf Sf} flashes occur with rate \eqref{rate} but are accompanied by no changes in the wave function.\footnote{Accordingly, equation \eqref{nflashdist} is replaced by
\begin{multline}
  \PPP\bigl( X_1\in d x_1, T_1 \in d t_1, I_1 = i_1, \ldots,
 X_n \in dx_n, T_n \in d t_n, I_n = i_n   |  \psi_{t_0} \bigr)  \\ 
=\lambda^n e^{-N\lambda (t_n-t_0)}   \prod_{k=1}^{n}
\langle \psi_{t_{k}} |  \Lambda_{i_k}(x_{k})\psi_{t_{k}}
\rangle \, dx_{1}dt_1   \cdots  dx_{n}dt_n \,, \nonumber
\end{multline}
where $\Lambda_{i}(x)$ is the collapse operator given by
\eqref{eq:collapseoperator}.}  (This flash process defines, in fact, a
Poisson process in space-time---more precisely, a Poisson system of points in
$\RRR^4\times\{1,\dots,,N\}$---with intensity measure
$r((x,t),i)=r(x,i|\psi_t)$ given by \eqref{rate}.) Note that, in contrast
to the case of {\sf GRWf}, one obtains a well defined theory by taking the
limit $\sigma\rightarrow 0$ in \eqref{eq:collapseoperator}, that is by
replacing $\Lambda_{i}(x)$ in \eqref{rate} with $\tilde{\Lambda}_i(x)$
given by $ \tilde{\Lambda}_i(x)=\delta(\widehat{Q}_i-x),
\label{eq:opsa}
$
where $\widehat{Q}_i$ is the position operator of the $i$-th `particle.'

Our last example ({\sf Sf$'$}) is the following.
Consider a nonrelativistic system of $N$ noninteracting quantum
particles with wave function satisfying the Schr\"odinger equation
\begin{equation}\label{schr}
  i\hbar \frac{\partial \psi}{\partial t} = -\sum_{i=1}^N
  \frac{\hbar^2}{2m_i} \nabla_i^2 \psi + \sum_{i=1}^N V_i(q_i)
  \, \psi\,,
\end{equation}
and suppose that, as in {\sf GRWf}, each of the flashes is associated with
one of the particle labels $1, \ldots, N$. 
Given the flashes up to the present, the next flash occurs with
rate $N\lambda$, and  has a label $I \in \{1, \ldots, N\}$ that is randomly chosen with uniform distribution. If this flash occurs at time $T_I$, 
its location $X$ is random with probability distribution
\begin{equation}\label{flashdistr}
  \mathbb{P}(X\in d X_I| I, T_I, \{X_k,T_k\}_{k\neq I}) = \mathcal{N} \, \bigl|\psi(X_1, T_1, \ldots, 
  X_N,T_N) \bigr|^2 \, d X_I \,, 
\end{equation}
where $\mathcal{N}$ is a normalizing factor, $\psi= \psi(q_1,t_1,
\ldots, q_N,t_N)$ is a multi-time wave function evolving according
to the set of $N$ equations
\begin{equation}\label{multitime}
  i\hbar \frac{\partial \psi}{\partial t_i} = -\frac{\hbar^2}{2m_i}
  \nabla_i^2 \psi + V_i(q_i) \, \psi
\end{equation}
for every $i \in \{1,\ldots, N\}$, and $T_k$ and $X_k$ are, for $k \neq I$,
the time and location of the last flash with label $k$. The reason that
this model is assumed to be noninteracting is precisely to guarantee the
existence of the multi-time wave function in \eqref{flashdistr}. {\sf Sf$'$}
is an example of a theory with a flash ontology that arguably is
empirically equivalent to {\sf OQT} (unlike {\sf GRWf})---at least, it
would be if it were extended to incorporate interactions between
particles---and avoids the many-worlds character of {\sf Sf} (see Section
\ref{subsec:swamw} below).

A provisional moral that emerges is that relativistic invariance might be
connected with a flash ontology, since {\sf GRWf} is the only theory in
Table 1 (except for {\sf Sm} and {\sf Sf}, which have a rather
extraordinary character that we discuss in Section \ref{subsec:swamw}
below) of which we know how it can be made relativistically invariant
without postulating a preferred foliation of space-time (or any other
equivalent additional structure). Finally, note that all the theories in
Table 1 are empirically equivalent (suitably understood) to {\sf OQT}
except {\sf GRWm}, {\sf GRWf}, and {\sf GRWp}.

\subsection{Schr\"odinger Wave Functions  and Many-Worlds}
\label{subsec:swamw}

A rather peculiar theory representing the world as if it were, at any given
 time, a collection of particles with classical configuration
 $Q=(Q_1,\ldots ,Q_N)$ is Bell's version of many-worlds ({\sf BMW})
 \citep{BellMW}. In {\sf BMW} the wave function $\psi$ evolves according to
  Schr\"odinger's equation and \citep{BellMW}
\begin{quote}
   instantaneous classical configurations \ldots{} are 
  supposed to exist, and to be distributed \ldots{}  with probability $|\psi|^2$.
  But no pairing of configurations at different times, as would be 
  effected by the existence of trajectories, is supposed.
\end{quote}
This can be understood as suggesting that the configurations at different times are not connected by any law. It could also be regarded as suggesting that configurations at different times are (statistically) independent, and that is how we shall understand it here.  
The world described by {\sf BMW} is so radically different from what we are accustomed to that it is hard to take {\sf BMW} seriously. In fact, for example, at some time during the past second, according to {\sf BMW}, there were on the earth dinosaurs instead of humans, because of the independence and the fact that, in any no-collapse version of quantum theory, there are parts of the wave function of the universe in which the dinosaurs have never become extinct. In this theory, the actual past will typically entirely disagree with what is suggested by our memories, by history books, by photographs and by other records of (what we call) the past. 

Also {\sf Sf} and {\sf Sm}, though they are  simple mathematical modifications of {\sf GRWf} and {\sf GRWm} respectively,  provide very different pictures of reality, so different indeed from 
 what we usually believe reality should be  like that it would seem hard to take these theories seriously. In {\sf Sf} and {\sf Sm}, apparatus pointers never  
point in a specific direction (except when a certain direction in {\sf OQT} would  
have probability more or less one), but rather all directions are, so to speak, realized at once.  As a consequence, one is led to conclude that their predictions don't agree with those of the quantum formalism. 
Still, it can be argued that these theories do not predict any \emph{observable} deviation from the quantum  
formalism: there is, arguably, no conceivable experiment that could  
help us decide whether our world is governed by {\sf Sf} or {\sf Sm} on the one hand or by the quantum formalism on the other. The reason for this surprising claim is that {\sf Sf} and {\sf Sm} can be regarded as many-worlds formulations of quantum mechanics. Let us explain. 

At first glance, in an {\sf Sf} or {\sf Sm} world, the after-measurement state of the  
apparatus seems only to suggest  that matter is very spread out.
However, if one considers the flashes, governed by the rate \eqref{rate}, or the mass density \eqref{mdef}, that correspond to  macroscopic superpositions, one sees that they form independent families 
of correlated flashes or mass density associated with the terms of the superposition, with no interaction between the families. The families can indeed be regarded as comprising many worlds, superimposed on a single space-time.  Metaphorically speaking, the universe according to {\sf Sf} or {\sf Sm} resembles the situation of a TV set that is not correctly tuned, so that one  always sees a mixture of two channels. In principle,  one might watch two movies at the same time in this way, with each movie conveying its own story composed of temporally and spatially correlated events.

Thus {\sf Sf} and {\sf Sm} are analogous to Everett's many-worlds ({\sf EMW}) formulation of quantum mechanics \citep{everett}, but with  the `worlds'  explicitly realized in the same space-time.
Since the different worlds do not interact among themselves---they are, so to speak, reciprocally transparent---this difference should not be regarded as crucial. Thus, to the extent that one is willing to grant that {\sf EMW} entails no observable deviation from the quantum  formalism, the same should be granted to {\sf Sf} and {\sf Sm}.  
Moreover, contrarily to {\sf EMW}, but similarly to {\sf BMW}, {\sf Sf} and
{\sf Sm} have a clear PO upon which the existence and behavior of the macroscopic counterparts of our experience  can be 
grounded. 

This ontological clarity notwithstanding, in {\sf Sf} and {\sf Sm} reality is of course very different from what we usually believe it  to be like. It is populated with ghosts we do not perceive, or rather, with what are like ghosts from our perspective, because the ghosts are as real as we are, and from their perspective we are the ghosts. We plan to give a more complete discussion of {\sf Sf} and {\sf Sm} in a future work.

We note that the theory {\sf Sm} is closely related to---if not precisely the same as---the version of quantum mechanics proposed by \citet{sch1}. After all, Schr\"odinger originally regarded his theory as describing a continuous distribution of matter (or charge) spread out in {\em physical space} in accord with the wave function on {\em configuration space} \citep{sch1}. He soon rejected this theory because he thought that it rather clearly conflicted with experiment.  Schr\"odinger's rejection of this theory was perhaps a bit hasty. Be that as it may, according to what we have said above, Schr\"odinger did in fact create the first many-worlds theory, though he probably was not aware that he had done so. (We wonder whether he would have been pleased if he had been).\footnote{However, Schr\"odinger did  write that \citep[p. 120]{sch} `$\psi\bar{\psi}$ is a
kind of weight-function in the system's configuration space. The
wave-mechanical configuration of the system is a superposition of
many, strictly speaking of all, point-mechanical configurations kinematically possible. Thus, each point-mechanical configuration contributes to the true wave-mechanical configuration with a certain
weight, which is given precisely by $\psi\bar{\psi}$. If we like paradoxes, we may
say that the system exists, as it were, simultaneously in all the
positions kinematically imaginable, but not  `equally strongly' in all.'}

\section{The Flexible Wave Function}
\label{sec:tfwf}
In this section we elaborate on the notion of physical  equivalence by
considering physically equivalent  formulations of {\sf{GRWf}} and {\sf{BM}}
for which the laws of evolution of the wave function are very different from
the standard ones. We conclude with some remarks on the notion of empirical
equivalence.

\subsection{GRWf Without Collapse}
\label{sec:lgrwf}
As a consequence of the view that the {\sf GRW} theory is ultimately not about wave functions but about either flashes or matter density, the
process $\psi_t$ in Hilbert space (representing the collapsing wave function) should no longer be regarded as playing the central role in
the {\sf GRW} theory. Instead, the central role is played by the random set $F$ of flashes for {\sf GRWf}, respectively by the random matter density function $m(\cdot,t)$ for {\sf GRWm}. 
From this understanding of {\sf GRWf} as being fundamentally about flashes, we obtain a lot of flexibility as to how we should regard the wave function and prescribe its behavior. As we point out in this section, it is not necessary to regard the wave function in {\sf GRWf} as undergoing collapse; instead, one can formulate {\sf GRWf} in such a way that it involves a wave function $\psi$  that evolves linearly (i.e., following the usual Schr\"odinger evolution).  

Suppose the wave function at time $t$ is $\psi_t$. Then according to equation \eqref{rate}, for {\sf{GRWf}} the rate for the next flash is given by 
\begin{equation}
r(x,i|\psi_t)=\lambda  
  \|\Lambda_i (x)^{1/2}\psi_{t}\|^2 .
  \label{ps}
\end{equation}
Observe that $\psi_t$, given by equation \eqref{eq:psit}, is determined by $\psi_{t_0}$ and the flashes $(X_k,T_k)$ that occur between the times $t_0$ and $t$; it 
can be rewritten as follows:
\begin{equation}\label{eq:newpsi}
  \psi_t =  
  \frac{ \Lambda_{I_n}(X_n,T_n;t)^{1/2} \;  \cdots\;   
  \Lambda_{I_1}(X_1,T_1;t)^{1/2}  \psi_{t}^L}
  {\|  \Lambda_{I_n}(X_n,T_n;t)^{1/2} \;  \cdots\;   
  \Lambda_{I_1}(X_1,T_1;t)^{1/2}  \psi_{t}^L\|}
\end{equation}
where we have introduced the  Heisenberg-evolved operators (with respect to time $t$)
\begin{equation}\Lambda_{I_k}(X_k,T_k;t)^{1/2} =  
U_{t-T_k} \Lambda_{I_k}(X_k)^{1/2} U_{T_k-t} = 
U_{t-T_k} \Lambda_{I_k}(X_k)^{1/2} U_{t-T_k}^{-1}
\end{equation}
and the linearly evolved wave function
\begin{equation}
  \psi_{t}^L= U_{t-t_0}\psi_{t_0}\, ,
\end{equation}     
where $t_0$ is the initial (universal) time.
By inserting $\psi_t$ given by equation \eqref{eq:newpsi} in \eqref{ps} one obtains that 
\begin{equation}
r(x,i|\psi_t)=\lambda \frac{  \| \Lambda_i (x)^{1/2}\Lambda_{I_n}(X_n,T_n;t)^{1/2} 
  \cdots\;   \Lambda_{I_1}(X_1,T_1;t)^{1/2}  \psi_{t}^L
\|^2}{\|   \Lambda_{I_n}(X_n,T_n;t)^{1/2} 
  \cdots\;   \Lambda_{I_1}(X_1,T_1;t)^{1/2}  \psi_{t}^L
  \|^2}.
\label{eq:newpsip}
\end{equation}

Suppose that the initial wave function is $\psi_{t_0}$, i.e., that the linearly evolved wave function at time $t$ is $\psi_t^L$. 
Then the right hand side of equation \eqref{eq:newpsip} defines the conditional rate for the next flash after time $t$, given the flashes in the past of $t$. 
Note that this conditional rate thus defines precisely the same flash process as {\sf GRWf}. In particular, we have that 
\begin{equation}
  \mathbb{P}_{\psi_t^L}( \mbox{future flashes} |  \mbox{past flashes}) = \mathbb{P}(\mbox{future flashes}   
  | \psi_t).
\label{ppp}
\end{equation}
The collapsed wave function $\psi_t$ provides precisely the same information as the linearly evolving wave function $\psi_t^L$ together with all the flashes.
Thus, one arrives at the  surprising conclusion  that the Schr\"odinger wave function 
 can be regarded as governing the evolution of the space-time point process of {\sf GRWf},
so that {\sf GRWf} can indeed be regarded as a  \textit{no-collapse}  theory involving flashes. We say `no-collapse' to underline that the dynamics of the PO is then governed by a wave function evolving according to the standard, linear Schr\"odinger equation \eqref{Schr}.  
However, while the probability distribution of the future flashes, given the collapsing wave function $\psi_t$, does not depend on the past flashes, given only $\psi_t^L$ it does. 

The two versions of {\sf GRWf}, one using the collapsing \wf $\psi_t$ and the other using the non--collapsing \wf $\psi_t^L$, should be regarded not as two different theories but rather as two formulations of the same theory, {\sf GRWf}, because they lead to the same distribution of the flashes and thus are physically equivalent.
We conclude from this discussion that what many have considered to be the crucial, irreducible difference between {\sf BM} and {\sf GRWf}, namely that the wave function collapses in {\sf GRWf} but does not in {\sf BM}, is not in fact an objective difference at all, but rather a matter of how {\sf GRWf} is presented. 
\bigskip
\label{sec:markov}

We close this section with a remark. A notable difference between the two presentations of {\sf GRWf} is that while the {\sf GRW} collapse process $\psi_t$ is a Markov process,\footnote{This means that
$\PPP\bigl(\text{future}\big|\text{past } \& \text{ present}\bigr) = \PPP\bigl(\text{future}\big|\text{present}\bigr)$. In
more detail, the distribution of the $\psi_t$ for all $t>t_0$ conditional on the $\psi_t$ for all $t \leq t_0$ coincides with the distribution of the future conditional on $\psi_{t_0}$.} 
the point-process $F$ of flashes is generically non--Markovian. In more detail, we regard a point process in space-time
as Markovian if for all $t_1<t_2$,
\begin{equation}
 \PPP\bigl(\text{future of }t_2\big|\text{past of }t_2\bigr)
  = \PPP\bigl(\text{future of }t_2 \big|
  \text{strip between $t_1$ and $t_2$}\bigr)\, ,
\end{equation}
where `future of $t_2$' refers to the configuration of points after time $t_2$, etc..
To see that $F$ is non--Markovian, note that the distribution
of the flashes in the future of $t_2$ depends on what happened between time
0 and time $t_2$, while the strip in space-time between $t_1$ and $t_2$ may provide
little or no useful information, as it may, for
small duration $t_2-t_1$, contain no flashes at all.\footnote{The matter density field $m(\cdot,t)$ is generically Markovian, but rather by coincidence: Given the initial wave function, different
patterns of collapse centers between time 0 and time $t_2$ should be expected to lead to different fields $m(\cdot,t_2)$, so
that the past (or equivalently $\psi_{t_2}$) may be mathematically determined from $m(\cdot,t_2)$.}

For a Markovian flash process events in a time interval $[t_1,t_2]$ are independent of those in a disjoint time interval $[t_3,t_4]$, which, as discussed in Section \ref{sec:differences}, would be rather unreasonable for a model of our world. In passing, we note that {\sf Sf} can indeed be regarded as a sort  
of  Markovian approximation of (the linear version of) {\sf GRWf} for  
which, at any time, the past is completely ignored in the  
computation of the conditional probability of future flashes.

\subsection{Bohmian Mechanics With Collapse}
\label{sec:BMC}
In Section \ref{sec:lgrwf} we showed that {\sf GRWf} can be reformulated in terms of a linearly evolving wave function. 
Conversely, {\sf BM} can be reformulated so that it involves a `collapsed' wave function. In this formulation the evolution of the wave function depends on the actual configuration. 
The state at time $t$ is described by the pair $(Q_t,\psi_t^C)$, where $Q=(Q_1,\ldots, Q_N)$ is the (usual) configuration but $\psi_t^C: \RRR^{3N} \to \CCC$
is a different wave function than usual, a collapsed wave function. Instead of equations \eqref{Bohm} and \eqref{Schr}, the state evolves according to
\begin{equation}\label{BohmC}
  \frac{dQ_i}{dt} = \frac{\hbar}{m_i} \Im \frac{\psi^{C*}\nabla_i \psi^C}{\psi^{C*}\psi^C} (Q_1,\ldots, Q_N)\,,
\end{equation}
which is the same as \eqref{Bohm} with $\psi$ replaced by $\psi^C$, and
\begin{equation}\label{SchrC}
  i\hbar\frac{\partial \psi^C}{\partial t} = - \sum_{i=1}^N
  \frac{\hbar^2}{2m_i} (\nabla_i- i\tilde A_i)^2 \psi^C +
  (V+\tilde V) \psi^C
\end{equation}
which is the same as  Schr\"odinger's equation except for the imaginary  pseudo-potentials ($\sigma \approx 10^{-7}$ m is the same constant as in {\sf GRW})
\begin{equation}
  \tilde A_i = \frac{i}{\sigma^2}(q_i -Q_i) \,, \quad \tilde V = -\frac{i}{\sigma^2}
  \sum_{i=1}^N \frac{\hbar^2}{m_i}(q_i-Q_i) \cdot \Im \frac{\psi^{C*} \nabla_i \psi^C}{\psi^{C*} \psi^C}\end{equation}
making equation \eqref{SchrC} nonlinear and $Q$-dependent. A solution $t \mapsto (Q_t, \psi_t^C)$ of equations \eqref{BohmC} and \eqref{SchrC} can be obtained from a solution $t \mapsto (Q_t,\psi_t)$ of equations \eqref{Bohm} and \eqref{Schr} by setting
\begin{equation}\label{C}
  \psi^C (q_1,\ldots,q_N) = \exp\biggl( - \sum_{i=1}^N
  \frac{(q_i-Q_i)^2}{2\sigma^2} \biggr) \, \psi (q_1,\ldots,q_N)\,.
\end{equation}
This is readily checked by inserting \eqref{C} into equations \eqref{BohmC} and \eqref{SchrC}. The ensemble of trajectories with distribution $|\psi|^2$ cannot be expressed in a simple way in terms of $\psi^C$. Nonetheless, for given initial configuration $Q_0$, we obtain from equations \eqref{BohmC} and \eqref{SchrC}, with given initial $\psi_0^C$, the same trajectory $t \mapsto Q_t$ as from equations \eqref{Bohm} and \eqref{Schr} with the corresponding $\psi_0$. This may be enough to speak of physical equivalence.

One can read off from \eqref{C} that $\psi^C$ is a \emph{collapsed} wave function: Whenever $\psi$ is a superposition (such as for Schr\"odinger's cat) of macroscopically different states with disjoint supports in configuration space, then in $\psi^C$ all contributions except the one containing the actual configuration $Q_t$ are damped down to near zero. (Still, the evolution is such that when two disjoint packets again overlap, the trajectories display an interference pattern.)

Of course, the unitarily-evolving $\psi_t$ is much more natural than $\psi_t^C$ as a mathematical tool for defining the trajectory $t\mapsto Q_t$; \eqref{Schr} is a simpler equation than \eqref{SchrC}. Still, the example shows that we have the choice in {\sf BM} between using a collapsed wave function $\psi^C$ or a spread out wave function $\psi$.

\subsection{Empirical Equivalence and Equivariance}
\label{sec:equiv}

The facts that {\sf GRWf} can be reformulated so that the wave function
evolves linearly, in the usual manner according to Schr\"odinger's
equation, and that {\sf BM} can be reformulated in terms of a collapsed
wave function indicate that the disagreement between the predictions of the
two theories should not be regarded as arising merely from the fact that
they involve different wave function evolutions. It is our contention that
the source of the empirical disagreement between {\sf BM} and {\sf GRWf}
can be regarded as lying, neither in their having different evolutions for the wave function,
nor in their having different ontologies, but rather in the presence or
absence of equivariance with respect to the Schr\"odinger evolution. More
explicitly, we claim that a theory is empirically equivalent to the quantum
formalism (i.e., that its predictions agree with those of the quantum
formalism) if it yields an equivariant distribution (defining typicality)
relative to the Schr\"odinger evolution that can be regarded as
`effectively $|\psi|^2$.'  Let us explain.

The view we have proposed about the PO of a theory and the corresponding role of the
wave function has immediate consequences for the criteria for the \emph{empirical
  equivalence} of two theories, i.e., the statement that they make
(exactly and always) the same predictions for the outcomes of experiments.

Before discussing these consequences, let us note a couple of remarkable
aspects of the notion of empirical equivalence. One is that, despite the
difficulty of formulating the \emph{empirical content} of a theory
precisely (a difficulty mainly owed to the vagueness of the notion
`macroscopic'), one can sometimes establish the empirical equivalence of
theories; for example, that of {\sf BM} and {\sf SM} or that of {\sf GRWm}
and {\sf GRWf}; for further examples see \citep{aapr}. Another remarkable
aspect is that empirical equivalence occurs at all. One might have expected
instead that different theories typically make different predictions, and
indeed the theories of classical physics would provide plenty examples. But
in quantum mechanics empirical equivalence is a widespread phenomenon; see
\citep{aapr} for discussion of this point.

Let us turn to the criteria for empirical equivalence. Since the empirical
equivalence of two theories basically amounts to the assertion that the two worlds,
governed by the two theories, share the same macroscopic appearance, we have to
focus on how to read off the macroscopic appearance of a possible world
according to a theory. And according to our view about PO, the macroscopic
appearance is a function of the PO---but not directly a function of the
wave function. In cases in which one can deduce the macroscopic appearance
of a system from its wave function, this is so only by virtue of a law of
the theory implying that this wave function is accompanied by a PO with a
certain macroscopic appearance. In short, empirical equivalence amounts to
a statement about the PO. This view is exemplified by our proof of empirical
equivalence between {\sf GRWm} and {\sf GRWf} in Section
\ref{sec:empequiv}. In more detail, the position $Z_t$ of, say, a pointer at time
(circa) $t$ is a function of the PO: In {\sf BM} and {\sf GRWm} it can be
regarded as a function $Z_t = Z(Q_t)$ of the configuration, respectively as
a function $Z_t = Z(m(\cdot,t))$ of the $m$ field, at time $t$, whereas in
{\sf GRWf} it is best regarded as a function of the history of flashes
over the past millisecond or so.

Concerning the empirical equivalence between a theory and {\sf OQT}, we need to
ask whether the probability of the event $Z_t=z$ agrees with the
distribution predicted by standard quantum mechanics.  The latter can be
obtained from the Schr\"odinger wave function $\psi_t$ for a sufficiently
big system containing the pointer by integrating $|\psi_t|^2$ over all
configurations in which the pointer points to $z$.  Thus, regardless of
what the PO of a theory is, all that is required for the empirical
equivalence between the theory and {\sf OQT} is that it provide the correct
$|\psi_t|^2$ probability distributions for the relevant variables $Z_t$.
When this is so we may speak of an `effective
$|\psi_t|^2$--distribution,' or of  \textit{macroscopic
  $|\psi|^2$ Schr\"odinger equivariance}. Thus, empirical equivalence to
{\sf OQT} amounts to having macroscopic $|\psi|^2$ Schr\"odinger
equivariance. (This applies to `normal' theories in which pointers point;
the situation is different for theories with a many-worlds character as
discussed in Section
\ref{subsec:swamw}.)

{\sf GRWf} (or {\sf GRWm}) predicts (approximately) the quantum mechanical
distribution only under certain circumstances, including, e.g., that the
experimental control over decoherence is limited, and that the universe is
young on the timescale of the `universal warming' predicted by {\sf
  GRWf}/{\sf GRWm} \citep[see][for details]{BG03}. Moreover, we know that
{\sf GRWf}, roughly speaking, makes the same predictions as does the
quantum formalism for short times, i.e., before too many collapses have
occurred. Thus, {\sf GRWf} yields an effective $|\psi|^2$--distribution for
times near the initial time $t_0$.  Now, if {\sf GRWf} were `effectively
$|\psi|^2$--equivariant,' its predictions would be the same as those of
quantum theory for all times. It is the absence of this macroscopic
$|\psi|^2$ Schr\"odinger equivariance that renders {\sf GRWf} empirically
inequivalent to quantum theory and to {\sf BM}. We shall elaborate on this
in a future work \citep{EEE}.

The most succinct expression of the source of the empirical disagreement
between {\sf BM} and {\sf GRWf} is thus the assertion that {\sf BM} is
effectively $|\psi|^2$-equivariant relative to the Schr\"odinger evolution
while {\sf GRWf} is not.  The macroscopic Schr\"odinger equivariance of
{\sf BM} follows, of course, from its microscopic $|\psi|^2$
Schr\"o\-din\-ger equivariance, while the lack of macroscopic $|\psi|^2$
Schr\"odinger equivariance for {\sf GRWf} follows from the warming
associated with the {\sf GRW} evolution and the fact that {\sf GRWf}, as
discussed in Section \ref{subsec:poae}, is microscopically equivariant
relative to that evolution. In fact, it follows from the {\sf GRW} warming
that there is, for {\sf GRWf}, no equivariant association $\psi \mapsto
\PPP^{\psi}$ with $\psi$ a Schr\"odinger-evolving wave
function.\footnote{Since the {\sf GRWf} flash process is non--Markovian, the
formulation of the notion of equivariant association given in Section
\ref{subsec:poae} is not appropriate here; instead, $\PPP^{\psi}$ should
now be understood to be a probability measure on the space $\Omega$ of
possible histories of the PO for all times, but one whose conditional
probabilities for the future of any time given its past are as prescribed,
here by the formula (\ref{ppp}). The association is \textit{equivariant} if
$T_{-t}^{\star}\PPP^{\psi}=\PPP^{\psi_t}$, with $T_\tau$ now the time
translation mapping on $\Omega$.}

\section{What is a Quantum Theory without Observers?}
\label{sec:essential}

To conclude, we delineate the common structure of
{\sf GRWm}, {\sf GRWf}, and {\sf BM}:

\begin{itemize}
\item[(i)] There is a clear primitive ontology, and it describes
  matter in space and time.
\item[(ii)] There is a state vector $\psi$ in Hilbert space that
  evolves either unitarily or, at least, for microscopic systems very  probably for a long time approximately unitarily. 
\item[(iii)] The state vector $\psi$ governs the behavior of the \textrm{PO} by
  means of (possibly stochastic) laws.
\item[(iv)] The theory provides a notion of a \emph{typical}
  history of the \textrm{PO} (of the universe), for example by a probability 
  distribution on the space of all possible histories; from this 
  notion of typicality the probabilistic predictions
  emerge.
\item[(v)] The predicted probability distribution of the macroscopic
  configuration at time $t$ determined by the \textrm{PO} (usually) agrees (at least approximately)
 with that of the quantum formalism.
\end{itemize}

The features (i)--(v) are common to these three theories, but they
are also desiderata, presumably even necessary conditions, for any
satisfactory quantum theory without observers.\footnote{A certain generalization of (i)--(v) is supported by the considerations in \citep{dm}, where it is argued that some systems in a Bohmian universe should be regarded as being governed, or guided, not by a vector $\psi$ in Hilbert space but by a density matrix $\rho$ on Hilbert space, the so-called conditional density matrix. But this does not amount to a big conceptual difference.}

\section*{Acknowledgments}
We thank Detlef D\"urr, Federico Laudisa and Mauro Dorato for helpful comments. 
S.~ Goldstein is supported in part by NSF Grant DMS-0504504. N. Zangh\`\i\ is supported in part by INFN.


\begin{thebibliography}{29}

\bibitem[Albert(1992)]{albert} Albert, D. Z. [1992]: \textit{Quantum Mechanics and Experience}, Cambridge, MA: Harvard University Press.

\bibitem[Albert(1996)]{albert1} Albert, D. Z. [1996]: `Elementary Quantum Metaphysics', 
	in J. Cushing, A. Fine and S. Goldstein (\textit{eds}), 1996,
	{\it {Bohmian Mechanics and Quantum Theory: An Appraisal}}, 
	Dordrecht: Kluwer, pp.~277--84.

\bibitem[Allori et al.(2005)Allori, Dorato, Laudisa and Zangh{\`\i}]{AZ05}
	Allori, V., Dorato, M., Laudisa, F. and Zangh\`\i, N. [2005]: 
	\textit{La natura delle cose, introduzione ai fondamenti e alla 
	filosofia della fisica}, Rome: Carocci.

\bibitem[Allori et al.(unpublished,a)Allori, D\"urr, Goldstein, Tumulka and Zangh{\`\i}]{EEE}
  Allori, V., D\"urr, D., Goldstein, S., Tumulka, R. and Zangh\`\i, N. [unpublished,a]:
  `Empirical Equivalence and Equivariance', in preparation.

\bibitem[Allori et al.(unpublished,b)Allori, Goldstein, Tumulka and Zangh{\`\i}]{semicolon}
  Allori, V., Goldstein, S., Tumulka, R. and Zangh\`\i, N. [unpublished,b]:
  `Semicolon and the Nature of Reality', in preparation.

\bibitem[Bargmann(1954)]{bargmann} Bargmann, V. [1954]: `On Unitary Ray Representations of Continuous Groups',   
\textit{Annals of Mathematics}, \textbf{ 59}, pp.~1--46.

\bibitem[Bassi and Ghirardi(2003)]{BG03} Bassi, A. and Ghirardi, G.C. [2003]:
 `Dynamical Reduction Models',
  \textit{Physics Reports}, \textbf{379}, 257--426. 
  
\bibitem[Bell(1966)]{Bell66} Bell, J. S. [1966]: `On the Problem of Hidden Variables in
  Quantum Mechanics', \textit{Reviews of Modern Physics}, \textbf{38}, pp.~447--52.
  Reprinted as chapter 1 of \citep{Bell87b}.

\bibitem[Bell(1976)]{Bell76} Bell, J. S. [1976]: `The Theory of Local Beables',
  \textit{Epistemological Letters}. Reprinted as chapter 7 of
  \citep{Bell87b}.

\bibitem[Bell(1980)]{Bell80} Bell, J. S. [1980]: `De Broglie--Bohm, Delayed-Choice
   Double-Slit Experiment, and Density Matrix',
   \textit{International Journal of Quantum Chemistry}, \textbf{14},
   pp.~155--59. Reprinted as
   chapter 14 of \citep{Bell87b}.

\bibitem[Bell(1981)]{BellMW} Bell, J. S. [1981]: `Quantum Mechanics for Cosmologists', in
  C.~Isham, R.~Penrose, and D.~Sciama (\textit{eds}), 1981, \textit{Quantum Gravity
    2}, Oxford: Clarendon Press, pp.~611--37. Reprinted as chapter 15 of
  \citep{Bell87b}.
    
\bibitem[Bell(1986)]{Bell86} Bell, J. S. [1986]: `Quantum field theory without observers',
  \textit{Physics Reports}, \textbf{137}, pp.~49--54. Reprinted 
  under the title `Beables for quantum field theory' as
  chapter 19 of \citep{Bell87b}.
  
\bibitem[Bell(1987a)]{Bell87} Bell, J. S. [1987a]: `Are There Quantum Jumps?', in C. W. Kilmister (\textit{ed.}), 1987,
  \textit{Schr\"odinger. Centenary Celebration of a Polymath}, Cambridge:
  Cambridge University Press, pp.~41--52. Reprinted as chapter 22 of
  \citep{Bell87b}.
  
\bibitem[Bell(1987b)]{Bell87b} Bell, J. S. [1987b]: \textit{Speakable and Unspeakable in
    Quantum Mechanics}. Cambridge: Cambridge University Press.

\bibitem[Bell(1989)]{Bell89} Bell, J. S. [1989]: `Toward An Exact Quantum
  Mechanics', in S.~Deser and R.~J.~Finkelstein (\textit{eds}), 1989,
  \textit{Themes in Contemporary Physics, II}, 
  Teaneck, NJ: World Scientific, pp.~1--26.

\bibitem[Bell(1990)]{Bell90} Bell, J. S. [1990]: `Against ``Measurement''', in A.I.~Miller (\textit{ed.}), 1990,
  \textit{Sixty-Two Years of Uncertainty: Historical, Philosophical,
    and Physical Inquiries into the Foundations of Quantum Physics}, 
    volume 226 of \textit{NATO ASI Series B}, New
  York: Plenum Press. Reprinted [1990] in \textit{Physics  World}
  \textbf{3(8)}, pp.~33--40.

\bibitem[Benatti et al.(1995)Benatti, Ghirardi and Grassi]{Ghi}
	Benatti, F., Ghirardi, G.C. and Grassi, R. [1995]: 
	`Describing the macroscopic world: closing the circle within
	the dynamical reduction program',
	\textit{Foundations of Physics}, {\bf 25}, pp.~5--38.

\bibitem[Berndl et al.(1995)Berndl, Daumer, D\"urr, Goldstein and Zangh{\`\i}]{survey}
  Berndl, K., Daumer, M., D\"urr, D., Goldstein, S. and Zangh\`\i, N. [1995]: 
  `A Survey of Bohmian Mechanics', \textit{Il Nuovo
  Cimento}, \textbf{110B}, pp.~737--50. 

\bibitem[Bohm(1952)]{Bohm52} Bohm, D. [1952]: `A Suggested Interpretation of the Quantum
  Theory in Terms of ``Hidden'' Variables, I and II', \textit{Physical
    Review}, \textbf{85}, pp.~166--93.

\bibitem[Bohm and Hiley(1993)]{BohmHiley} Bohm, D. and Hiley, B.J. [1993]: \textit{The Undivided Universe}, London: Routledge.  

\bibitem[Dowker and Henson(2004)]{Fay02} Dowker, F. and Henson, J. [2004]: `Spontaneous Collapse Models on a
  Lattice', \textit{Journal of Statistical Physics} \textbf{115}, pp.~1327--39.

\bibitem[Dowker and Herbauts(2004)]{Fay03} Dowker, F. and Herbauts, I. [2004]: `Simulating Causal Wave-Function Collapse Models', \textit{Classical and Quantum Gravity}, \textbf{21}, pp.~1--17. 

 \bibitem[Dowker and Herbauts(2005)]{Fay04} Dowker, F. and Herbauts, I. [2005]: `The Status of the Wave Function in Dynamical
Collapse Models', \textit{Foundations of Physics Letters}, \textbf{18}, pp.~499--518. 

\bibitem[D\"urr et al.(2004a)D\"urr, Goldstein, Tumulka and Zangh{\`\i}]{crlet}
  D{\"u}rr, D., Goldstein, S., Tumulka, R. and
  Zangh{\`{\i}}, N. [2004a]: `Bohmian Mechanics and Quantum Field Theory',
  \textit{Physical Review Letters}, \textbf{93}, p.~090402.

\bibitem[D\"urr et al.(2005a)D\"urr, Goldstein, Tumulka and Zangh{\`\i}]{dm} D{\"u}rr, D., Goldstein, S., Tumulka, R. and
  Zangh{\`{\i}}, N. [2005a]: `On the Role of Density Matrices in Bohmian Mechanics',
  \textit{Foundations of Physics}, \textbf{35}, pp.~449--67. 

\bibitem[D\"urr et al.(2005b)D\"urr, Goldstein, Tumulka and Zangh{\`\i}]{crea2B} D{\"u}rr, D., Goldstein, S., Tumulka, R. and
  Zangh{\`{\i}}, N. [2005b]: `Bell-Type Quantum Field Theories',
\textit{Journal of Physics A: Mathematical and General}, \textbf{ 38}, pp.~R1--R43.

\bibitem[D\"urr et al.(1992)D\"urr, Goldstein and Zangh{\`\i}]{DGZ92} D\"urr, D., Goldstein, S. and Zangh\`\i, N. [1992]: `Quantum
  Equilibrium and the Origin of Absolute Uncertainty', \textit{Journal of
  Statistical Physics}, \textbf{67}, pp.~843--907. 

\bibitem[D\"urr et al.(2004b)D\"urr, Goldstein and Zangh{\`\i}]{DGZ04} D\"urr, D., Goldstein, S. and Zangh\`\i, N. [2004b]: `Quantum
  Equilibrium and the Role of Operators as Observables in Quantum 
  Theory', \textit{Journal of Statistical Physics}, \textbf{116}, pp.~959--1055.

\bibitem[Everett(1957)]{everett} Everett, H. [1957]:  `Relative State Formulation of Quantum Mechanics',
\textit{Reviews of Modern Physics}, \textbf{29}, pp.~454--62.

\bibitem[Ghirardi et al.(1986)Ghirardi, Rimini and Weber]{GRW86} Ghirardi, G.C., Rimini, A., Weber, T. [1986]: `Unified
  Dynamics for Microscopic and Macroscopic Systems', \textit{Physical Review
    D}, \textbf{34}, pp.~470--91.

\bibitem[Goldstein(1987)]{stochmech2} Goldstein, S. [1987]: `Stochastic Mechanics and Quantum
  Theory', \textit{Journal of Statistical Physics}, \textbf{47}, pp.~645--67.
  
\bibitem[Goldstein(1998)]{Gol98} Goldstein, S. [1998]: `Quantum Theory Without Observers',
  \textit{Physics Today}, Part One: March, pp.~42--6; 
  Part Two: April, pp.~38--42.
 
\bibitem[Goldstein et al.(2005)Goldstein, Taylor, Tumulka and Zangh{\`\i}]{aapr}
  Goldstein, S., Taylor, J., Tumulka, R. and Zangh\`\i, N. [2005]: 
  `Are All Particles Real?',
  \textit{Studies in History and Philosophy of Modern Physics},
  \textbf{36}, pp.~103--12.  

\bibitem[Kent(1989)]{kent} Kent, A. [1989]: `{}``Quantum Jumps'' and Indistinguishability',
  \textit{Modern Physics Letters A}, \textbf{4(19)}, pp.~1839--45.

\bibitem[Lewis(2005)]{plewis} Lewis, P. [2005]: `Interpreting Spontaneous Collapse Theories', {\it Studies in History and Philosophy of Modern Physics}, \textbf{36}, pp.~165--80.

\bibitem[Maudlin(forthcoming)]{Mau05} Maudlin, T. [forthcoming]: `Non-Local Correlations in Quantum Theory: Some Ways the Trick Might Be Done', to appear in Q.~Smith and   W.~L.~Craig (\textit{eds}),
  \textit{Einstein, Relativity, and Absolute Simultaneity},
  London: Routledge.

\bibitem[Monton(2002)]{monton} Monton, B. [2002]: `Wave Function Ontology',
	\textit{Synthese}, \textbf{130}, pp.~265--77.
	
\bibitem[Nelson(1985)]{stochmech1} Nelson, E. [1985]: \textit{Quantum Fluctuations},
  Princeton: Princeton University Press.

\bibitem[Nicrosini and Rimini(2003)]{rimini} Nicrosini, O. and Rimini, A. [2003]:
  `Relativistic Spontaneous Localization: a Proposal', 
  {\it Foundations of Physics}, {\bf 33}, pp.~1061--84. 

\bibitem[Pearle(1976)]{Pe76}  Pearle, P. [1976]: `Reduction of the State Vector by a 
   Nonlinear Schr\"odinger equation', \textit{Physical Review D}, \textbf{13},  
   pp.~857--68.

\bibitem[Pearle and Squires(1994)]{PS94} Pearle, P. and Squires, E. [1994]: `Bound State Excitation, Nucleon 
   Decay Experiments and Models of Wave Function Collapse',
   \textit{Physical Review Letters}, \textbf{73}, pp.~1--5.
   
\bibitem[Putnam(2005)]{putnam} Putnam, H. [2005]:
	`A Philosopher Looks at Quantum Mechanics (Again)',
   \textit{British Journal for the Philosophy of Science}, \textbf{56}, pp.~615--34.

\bibitem[Schr\"odinger(1926)]{sch1}Schr\"odinger, E. [1926]: `Quantisierung als Eigenwertproblem (Vierte Mitteilung)',
\textit{Annalen der Physik}, \textbf{81}, pp.~109--39.  English translation in \citep{sch}.

\bibitem[Schr\"odinger(1927)]{sch} Schr\"odinger, E. [1927]: \textit{Collected Papers on Wave Mechanics}, translated by J. F. Shearer, New York: Chelsea.

\bibitem[Struyve and Westman(2006)]{westman} Struyve, W. and Westman, H. [2006]:
  `A New Pilot-Wave Model for Quantum Field Theory',
  in A. Bassi, D. D\"urr, T. Weber and N. Zangh{\`{\i}} (\textit{eds}),
  \textit{Quantum Mechanics: Are there Quantum Jumps? and
  On the Present Status of Quantum Mechanics},
  AIP Conference Proceedings, \textbf{844},
  American Institute of Physics,  pp.~321--39.

\bibitem[Wigner(1939)]{wigner} Wigner, E.P. [1939]: `On Unitary Representations of the Inhomogeneous Lorentz Group', \textit{Annals of Mathematics}, {\bf 40},
pp.~149--204.

\bibitem[Tumulka(2006a)]{Tum04} Tumulka, R. [2006a]: `A Relativistic Version of the
  Ghirardi--Rimini--Weber Model',
  \textit{Journal of Statistical Physics}, \textbf{125}, pp.~821--40.

\bibitem[Tumulka(2006b)]{Tum05} Tumulka, R. [2006b]: `On Spontaneous Wave Function
	Collapse and Quantum Field Theory',
	\textit{Proceedings of the Royal Society A},
	\textbf{462}, pp.~1897--908.
	
\bibitem[Tumulka(2006c)]{Tum06} Tumulka, R. [2006c]:
  `Collapse and Relativity', in A.
  Bassi, D. D\"urr, T. Weber and N. Zangh{\`{\i}} (\textit{eds}),
  \textit{Quantum Mechanics: Are there Quantum Jumps? and
  On the Present Status of Quantum Mechanics},
  AIP Conference Proceedings, \textbf{844}, 
  American Institute of Physics, pp.~340--52.

\end{thebibliography}
\end{document}